\documentclass[twocolumn,twocolappendix]{aastex631}
\pdfoutput=1

\newcommand{\sgra}{Sgr~A$^*$}

\usepackage{graphicx} 
\usepackage{amsmath}
\usepackage{subfigure}
\usepackage{amssymb}

\begin{document}

\title{Modeling Targets and Optimal Frequencies for Imaging the Shadows of Nearby Supermassive Black Holes}

\author[0000-0003-4440-8924]{J. Cole Faggert}
\affiliation{School of Physics, Georgia Institute of Technology, 837 State St NW, Atlanta, GA 30332, USA}

\author[0000-0003-4413-1523]{Feryal \"Ozel}
\affiliation{School of Physics, Georgia Institute of Technology, 837 State St NW, Atlanta, GA 30332, USA}

\author[0000-0003-1035-3240]{Dimitrios Psaltis}
\affiliation{School of Physics, Georgia Institute of Technology, 837 State St NW, Atlanta, GA 30332, USA}

\begin{abstract}
Horizon-scale imaging of supermassive black holes has opened a new window onto the studies of strong-field gravity and plasma physics in low-luminosity accretion flows. As future efforts aim to image fainter and smaller angular-size targets, primarily through space-based very long baseline interferometry (VLBI), it is important to identify optimal sources and observing strategies for such studies. In this work, we assess the prospects for imaging black hole shadows in a broad population of nearby supermassive black holes by modeling their accretion flows using a covariant semi-analytic model for the flow and general relativistic ray tracing. We explore the influence of black hole and accretion flow parameters on spectra, image morphology, and the critical frequency at which the flows become optically thin. We identify three general classes of sources: those that become transparent at traditional imaging frequencies; those requiring higher frequencies; and those unlikely to be transparent down to the black hole shadow in the submillimeter band. Our results will inform target selection and wavelength optimization for future VLBI arrays, where both resolution and transparency are essential for resolving black hole shadows.
\end{abstract}

\keywords{Supermassive black holes (1663) --- Accretion (14) --- Very long baseline interferometry (1769)}

\section{Introduction} \label{sec:intro}

The horizon-scale images of the two black holes M87 and Sagittarius A* (\sgra) have opened a new window onto understanding the rich gravitational physics of black holes and the plasma physics of accretion flows \citep{M87PaperI,SgrAPaperI}. These observations provided evidence for black hole shadows, direct tests of strong field gravity, and insights into the magnetic fields and thermodynamics of the surrounding plasmas \citep{M87PaperVI,SgrAPaperVI,M87PaperVIII}. 

\begin{deluxetable*}{lccccccc}[t]
\tablecolumns{6 }
\tablecaption{Black Hole Targets\label{tab:candidates}}
\tablehead{
\colhead{Name} & \colhead{Flux (Jy)} & \colhead{Mass ($10^8~M_\odot$)} & \colhead{Distance (Mpc)} & 
\colhead{Shadow Diameter ($\mu$as)} & \colhead{Estimated $i$}
}
\startdata
\sgra\ & $2.4$ & $0.043$ & $0.0083$ & $53.14$ & \\
M87 & $0.5$ & $62$ & $16.8$ & $37.85$&  \\
M60$^\dag$ & $1.7\times 10^{-3}$ & $45^{+10}_{-10}$ & $17.38$ & $26.56$&  \\
IC 1459 & $0.217$ & $26^{+11}_{-11}$ & $26.18$ & $10.19$&  \\
NGC 3115$^\dag$ & $\le 2\times 10^{-4}$ & $9.6^{+0.9}_{-1.9}$ & $10.3$ & $9.56$&  \\
NGC 4594 & $0.198$ & $6.6^{+0.4}_{-0.4}$ & $9.55$ & $7.09$& $66^{+4}_{-6} \;^\circ$ \\
NGC 3998 & $0.13$ & $8.1^{+2}_{-1.9}$ & $14.2$ & $5.85$&  \\
NGC 2663 & $0.084$ & $16$ & $28.5$ & $5.76$ &  \\
NGC 4261 & $0.2$ & $16.7^{+4.9}_{-3.4}$ & $32.4$ & $5.29$  & $54^\circ-84^\circ$ \\
M84 & $0.133$ & $8.5^{+0.9}_{-0.8}$ & $17$ & $5.13$&  $58^{+17}_{-18} \;^\circ$\\
NGC 3894 & $0.058$ & $20$ & $48.2$ & $4.26$&  $10^\circ-20^\circ$\\
3C 317 & $0.034$ & $46^{+3}_{-3}$& $122$ & $3.85$ & $81^\circ-85^\circ$\\
NGC 4552 & $0.027$ & $4.9^{+0.7}_{-0.4}$ & $16.3$ & $3.08$ & \\
NGC 315 & $0.182$ & $20.8^{+3.3}_{-1.4}$ & $70$ & $3.05$ & $\sim 50^\circ$\\
NGC 1218 & $0.11$ & $31.6$ & $116$ & $2.8$&  $\sim 20^\circ$\\
IC 4296 & $0.212$ & $13.4^{+2.1}_{-1.9}$ & $50.8$ & $2.71$&  \\
NGC 1399 & $0.038$ & $5.1^{+0.7}_{-0.7}$ & $21.1$ & $2.48$ & \\
Cen A & $5.98$ & $0.55^{+0.3}_{-0.3}$ & $3.42$ & $1.65$ & \\
\enddata
\tablerefs{See Appendix \ref{sec:Appendix A} for further detail and references for each of the sources. The two sources marked with a dagger are significantly fainter than the rest.}
\vspace*{-0.5cm}
\end{deluxetable*}

\vspace*{-0.8cm}
As we aim to broaden horizon-scale imaging to more sources, which are smaller in angular size than the first targets, it becomes necessary to increase angular resolution either by extending interferometric baselines to space, decreasing the wavelength of observations, or both. There have indeed been such efforts and proposals, starting with EHT observations at smaller wavelengths  (corresponding to 345~GHz; \citealt{EHT345GHz}), as well as concepts to carry out millimeter-wave interferometry with one or more space baselines \citep{EHimager,Millimetron,Capella,BHEX_Motivation,ChineseSpaceMillimeterVLBI,zineb2024advancing}. An important step for all of these endeavors  is to identify optimal targets that are resolvable, bright, and are transparent down to their horizons to enable imaging of their shadows. The first of these factors is determined by the mass-to-distance ratio $M/D$ for each target. The other two depend on the accretion rate and the wavelength of observation. 

Nearby supermassive black holes, especially those that have large angular sizes and bright millimeter emission, belong to a class of low-luminosity, radiatively inefficient accretors \citep{Yuan2014}. Radio emission in such accretion flows is dominated by synchrotron emission from hot electrons. At long wavelengths, the synchrotron emission is self-absorbed and the accretion flow is opaque. As the frequency increases, the opacity rapidly decreases and, at some critical frequency, the accretion flow becomes optically thin down to the horizon. Notably, this transition frequency is also near the frequency at which the source is at its brightest at that band. 

The transition frequency is, in principle, different for each source as it is affected by a variety of physical properties of the black hole and of the accretion flow.  In particular, the transition is affected primarily by the black hole mass and the mass accretion rate \citep{OzelOpticallyThin}. Increasing the mass accretion rate increases the density scale, which makes the accretion flow more opaque and pushes the transition toward higher frequencies. On the other hand, increasing the mass of the black hole requires less density to achieve the same brightness, moving the transition frequency to lower values. For the EHT imaging targets, these considerations were addressed by means of theoretical models. Remarkably, for both M87 and \sgra, which have widely different properties, this transition happens at $\sim $mm wavelengths for a broad range of flow characteristics, which is what enabled the EHT to observe their shadows \citep{OzelOpticallyThin}.

In this paper, we identify a set of nearby targets based on their millimeter-wave brightness and shadow diameters and model their spectra with a wide range of theoretical models to determine the optimal frequency of observation. We base our initial selection on the promising candidates identified by \citet{Johannsen2012} and supplement the target list and observations with the recent ETHER catalog \citep{ETHER}, as well as with Very Large Array (VLA), Very Long Baseline Array (VLBA), and Atacama Large Millimeter-Submillimeter Array (ALMA) observations, where available. We use a covariant semi-analytic model for the accretion flow~\citep{BlackHoleImagesAsTestPlasma}, which enables us to explore the dependence of spectra and opacities on the density, temperature, and magnetic field profiles in the accretion flows. We also explore the impact of plasma parameters such as the ion-to-electron temperature ratios and magnetic field equipartition. 

Our approach differs from that of the recent study by \citet{ngEHTsource+modelling} in two aspects. First, instead of assuming a single model (in fact, a single snapshot) for the accretion flow from a numerical simulation, with the same black hole spin and plasma parameters, we perform a systematic exploration of black hole, accretion flow, and observer parameters. Second, we explore a broad range of observing wavelengths to identify the optimal one that will allow unobstructed imaging of the black-hole shadows.

\begin{figure}[t]
\includegraphics[width=0.99\linewidth]{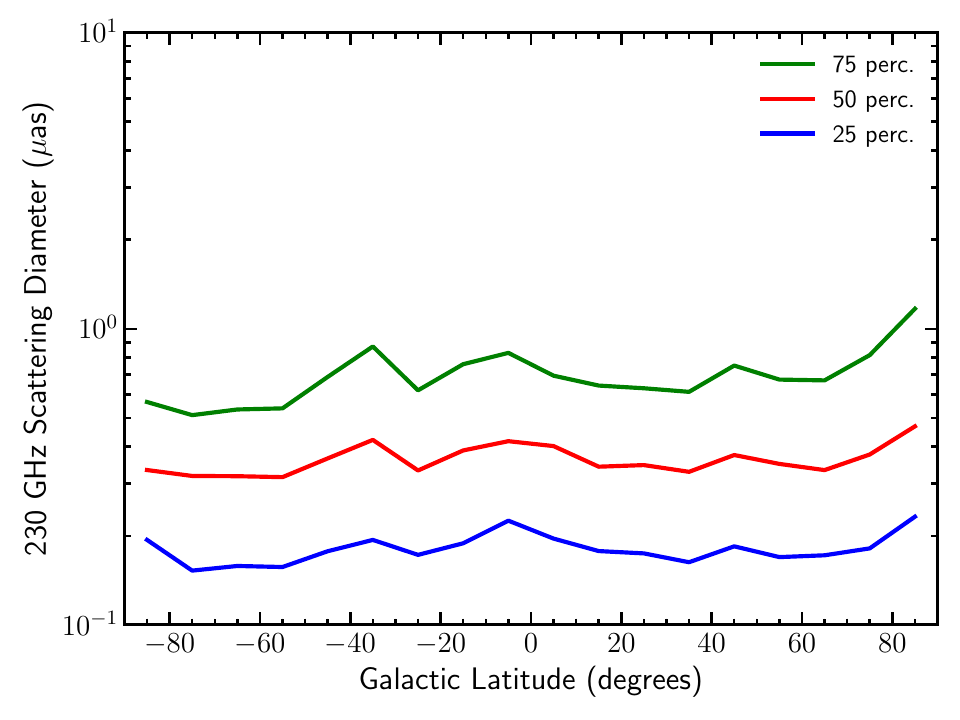}
\caption{Distribution of predicted angular diameters of scatter-broadened point-source images at 230~GHz as a function of Galactic latitude. The green, red, and blue curves represent the 75th, 50th, and 25th percentiles, respectively, derived from AGN broadening measurements at lower frequencies. To ensure resolvability, we require black hole shadow diameters to be at least a factor of three larger than the 75-th percentile diameter, leading to a minimum shadow diameter of 1.5~$\mu$as.}
\label{fig:scattering}
\end{figure}

We begin in \S\ref{sec:sources} by identifying promising supermassive black hole targets based upon their angular size in the sky and their millimeter-band flux. In \S\ref{sec:model}, we introduce the accretion flow and emission models. In \S \ref{sec:properties}, we infer the mass accretion rates of the targets and characterize the images. Then, in \S\ref{sec:spectra}, we separate the targets into distinct categories based on their spectral properties and identify the optimal wavelengths for future imaging instruments. We present our conclusions in \S\ref{sec:conclusions}.

\section{Selecting Black Hole Targets} 
\label{sec:sources}

We first identify nearby supermassive black holes that have the largest known angular diameters of their shadows and have been detected as millimeter sources. To that end, we start from the study by \citet{JohannsenBlackHoles}, who identified promising, nearby candidates for space-based VLBI observations based on these two criteria. We augment this with sources and additional observations from the ETHER survey \citep{ETHER}. In addition, we have collected measurements at a broad range of radio frequencies from VLA, VLBA, and ALMA, which we discuss in detail in Appendix~\ref{sec:Appendix A} for each source. We list the resulting targets and their main characteristics in Table~\ref{tab:candidates}. When available, we have also included the observer's inclination, which was estimated from the orientation of their jets with respect to the line-of-sight. 

We introduce a lower limit on the angular diameter of the black-hole shadow based on considerations related to the expected broadening of the images caused by interstellar scattering. The presence of free electrons in our galaxy introduces diffractive broadening of images at a level that scales with the square of the wavelength of observations. \cite{Koryukova2022} used archival VLBI data of the cores of 8959 active galactic nuclei (AGN) at different orientations in the sky, observed at a range of wavelengths. We used these measurements of scatter broadening to predict the diameters of scattered point sources at 230~GHz as a function of galactic latitude. Figure~\ref{fig:scattering} shows the 75-th, 50-th, and 25-th percentiles for these distributions, with the former being $\sim 0.5\mu$as at all latitudes. Requiring that the angular diameters of black-hole shadows are at least 3 times larger results in a minimum shadow diameter $D_{\text{shadow}} = 1.5\;\mu$as, which we adopt as our threshold.

\begin{figure}[t]
\includegraphics[width=0.99\linewidth]{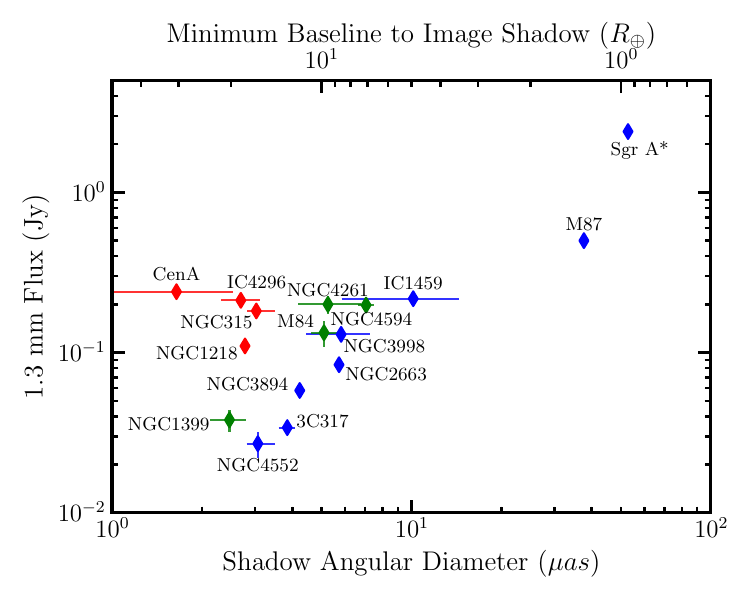}
\caption{1.3~mm fluxes and the shadow angular diameters of nearby supermassive black hole targets. Top axis shows the minimum baseline length necessary to image the black-hole shadows in these sources. The colors of the data point correspond to those in Figure~\ref{fig:transitionfrequency}.}
\label{fig:bhsources}
\end{figure}

Among these targets, two are exceptionally faint, with 1.3~mm fluxes of $\sim 1.7\times 10^{-3}$~Jy for M60 and $\lesssim 2\times 10^{-4}$~Jy for NGC\;3115, limiting their observability by interferometers. In our analysis, we focus on the remaining sources, which all have fluxes that exceed 10~mJy. 

Figure~\ref{fig:bhsources} shows the 1.3~mm fluxes and the shadow angular diameters for the sources listed in Table~\ref{tab:candidates}. 
In addition, the top axis shows the minimum baseline required to detect the presence of a brightness depression (shadow) in a  black hole image, using Eq.~(11) of \citet{zineb2024advancing}
\begin{equation} \label{eq: minbaseline}
    b_2 \simeq \frac{1.75}{\theta_{\text{shadow}}} \lambda.
\end{equation}

In \S\ref{sec:properties}, we will use the existing observations of these targets to assess the observability of their event horizons as a function of observing frequency and determine the optimal frequency of observation to reach their shadows. 

\section{Theoretical Modeling}
\label{sec:model}

The radio spectra and images of radiatively inefficient black holes are shaped by synchrotron emission, which depends on the magnetic field strength, electron temperature, and density of the plasma throughout the accretion flow. Our goal is to enable modeling of the images and spectra without imposing strong assumptions about the dissipation mechanism of angular momentum or the thermodynamics of the accretion flow. In order to explore the observability of the shadow under a broad range of plausible conditions, we use the covariant, semi-analytic model for radiatively inefficient flows described in \citet{BlackHoleImagesAsTestPlasma} and \citet{BlackHoleImagesAsTestSpacetime}. 

The covariant model solves for the density $\rho$ using the equation for conservation of mass,
\begin{equation}
T^{\mu\nu}_{;\nu}=0\;
\end{equation}
for the ion temperature using the conservation of energy-momentum, 
\begin{equation}
    (\rho u^\mu)_{;\nu}=0\;,
\end{equation}
and for the azimuthal plasma velocity following the assumption that gravity is the dominant force in the system. Here $T^{\mu\nu}$ and $u^\mu$ are the stress-energy tensor and 4-velocity of the plasma, respectively.

Because the dissipation of angular momentum due to turbulent stresses in the accretion flow affects the equatorial radial velocity profile, the model parametrizes the latter as
\begin{equation} 
\label{eq:radial velocity profile}
    u^r_{eq}(r)=-\eta \left(\frac{r}{r_{\rm ISCO}}\right)^{-n_r}
\end{equation}
in terms of a power-law index $n_r$ and an amplitude $\eta$. This is equivalent to parametrizing the unknown efficiency of angular momentum transport. Here, $r_{\rm ISCO}$ is the radius of the innermost circular stable orbit, inside which the material is assumed to plunge into the black hole. The model also allows us to vary the ratio of gas pressure $P$ to magnetic pressure, via the plasma $\beta$ defined as 
\begin{equation}
    \beta \equiv \frac{P}{B^2/8\pi}\;.
\end{equation}
This controls the strength $B$ of the magnetic field and directly affects the emissivity of synchrotron radiation in the flow. 

\begin{deluxetable}{cc} 
\tablewidth{4cm}
\tablecaption{Black Hole and Accretion Model Parameters 
\label{tab:parameters}}
\tablehead{
    \colhead{Parameter} & \colhead{Values} 
}
\startdata
$n_r$ & 1.0, 1.5 \\
$\beta$ & 5, 10  \\
$\frac{\zeta}{R+1}$ & 0.25/6, 0.4/6\\
$a$ & 0.1, 0.3, 0.5, 0.7, 0.9 \\
$i$ & $15^\circ$, $30^\circ$,  $45^\circ$,  $60^\circ$ \\
\enddata
\end{deluxetable}
\vspace*{-0.4cm}

Finally, the model allows us to to explore different scenarios for the electron temperature via the parametrization
\begin{equation}
    \frac{kT_e}{mc^2} = (\gamma -1) \frac{\zeta}{R+1} \left( \frac{GM}{rc^2}\right)\;, 
    \label{eq:Te}
\end{equation}
where $\gamma$ is the adiabatic index, $R$ is the   electron-to-ion temperature ratio, $\zeta$ is the ratio of the ion temperature to its virial value, $M$ is the mass of the black hole, $G$ is the gravitational constant, and $c$ is the speed of light. (See~\citealt{BlackHoleImagesAsTestPlasma} for details.) We choose a value for the ion-to-electron temperature ratio of $R=5$ (see \citealt{Satapathy2023,Satapathy2024}) and values for the fraction of the ion virial temperature in the range $\zeta= 0.25 - 0.4$, which are consistent with results of GRMHD simulations \citep{BlackHoleImagesAsTestPlasma}.

After solving for the properties of the accretion flow, we calculate the thermal synchrotron emissivity using the analytic fitting formula from \citet{Synchrotron}. We perform relativistic ray-tracing to integrate the radiative transfer equation along geodesics from the observer through the volume of the accretion flow. To perform these radiative transfer calculations, we have developed the algorithm \texttt{KRAY}, which utilizes the approach introduced in \citet{PsaltisRay} to perform backward ray-tracing from the image plane to the source. In \texttt{KRAY}, we parallelized the numerical solution through implementation in \texttt{KOKKOS}, allowing for easy use on single-core CPU, multi-core CPUs, and GPUs. As we are interested in the multi-frequency properties of the targets, and especially in the frequency of transition from optically thick to optically thin emission, we have also included synchrotron self-absorption in the integration of the transfer equation.

Using this model, we calculate images and spectra that span a wide range of accretion flow and black hole properties. 
For the properties of the black hole, we vary the black hole spin, $a$, as well as the observer's viewing inclination, $i$, with respect to the spin axis. We show the range and intervals of each of these parameters in Table~\ref{tab:parameters}.

The mass accretion rate sets the density scale in the accretion flow and is expressed in terms of $\dot{M}_{\rm Edd} = 4\pi G M m_p/\sigma_T \epsilon c$. Here $m_p$ is the proton mass, $\sigma_T$ is the Thomson scattering cross section, and $\epsilon=0.1$ is a fiducial radiative efficiency used only in this definition. In the next section, we will infer the accretion rate for each target by matching model predictions to the observed 230~GHz fluxes.

\begin{figure}[t]
    \centering
    \includegraphics[width=0.99\linewidth]{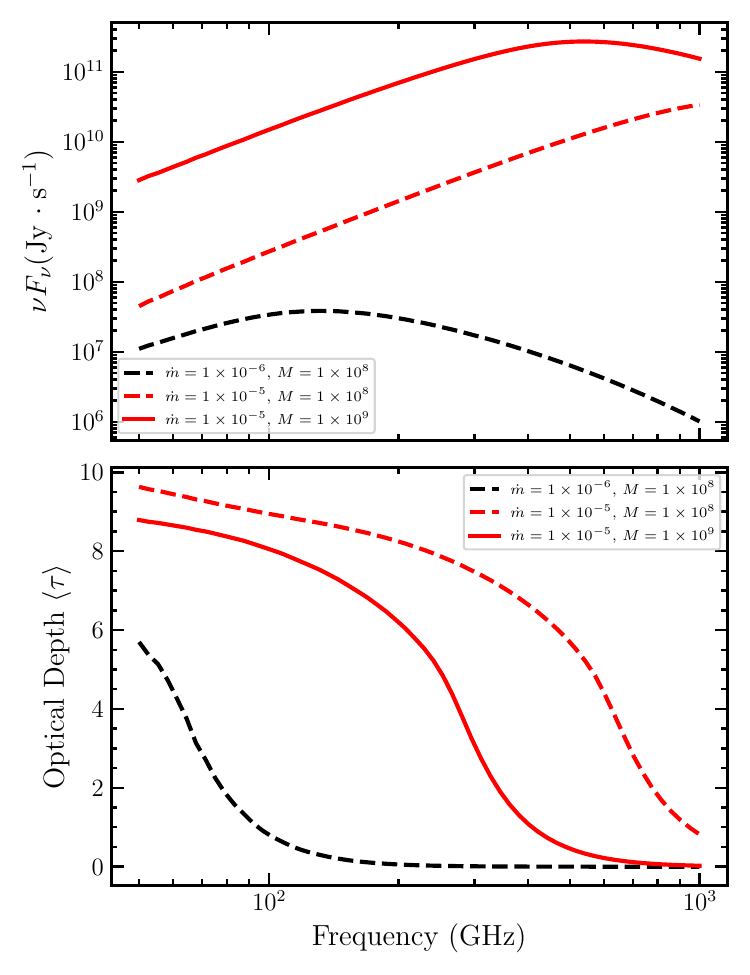}
    \caption{{\em (Top)\/} Effect of changing the black hole mass and mass accretion rate on the spectrum. {\em (Bottom)\/} The optical depth averaged inside the boundary of the black hole shadow as a function of frequency for the three configurations shown in the the top panel.}
    \label{fig:example}
\end{figure}

\begin{figure*}[ht!]
\centering
\includegraphics[width=1.0\columnwidth]{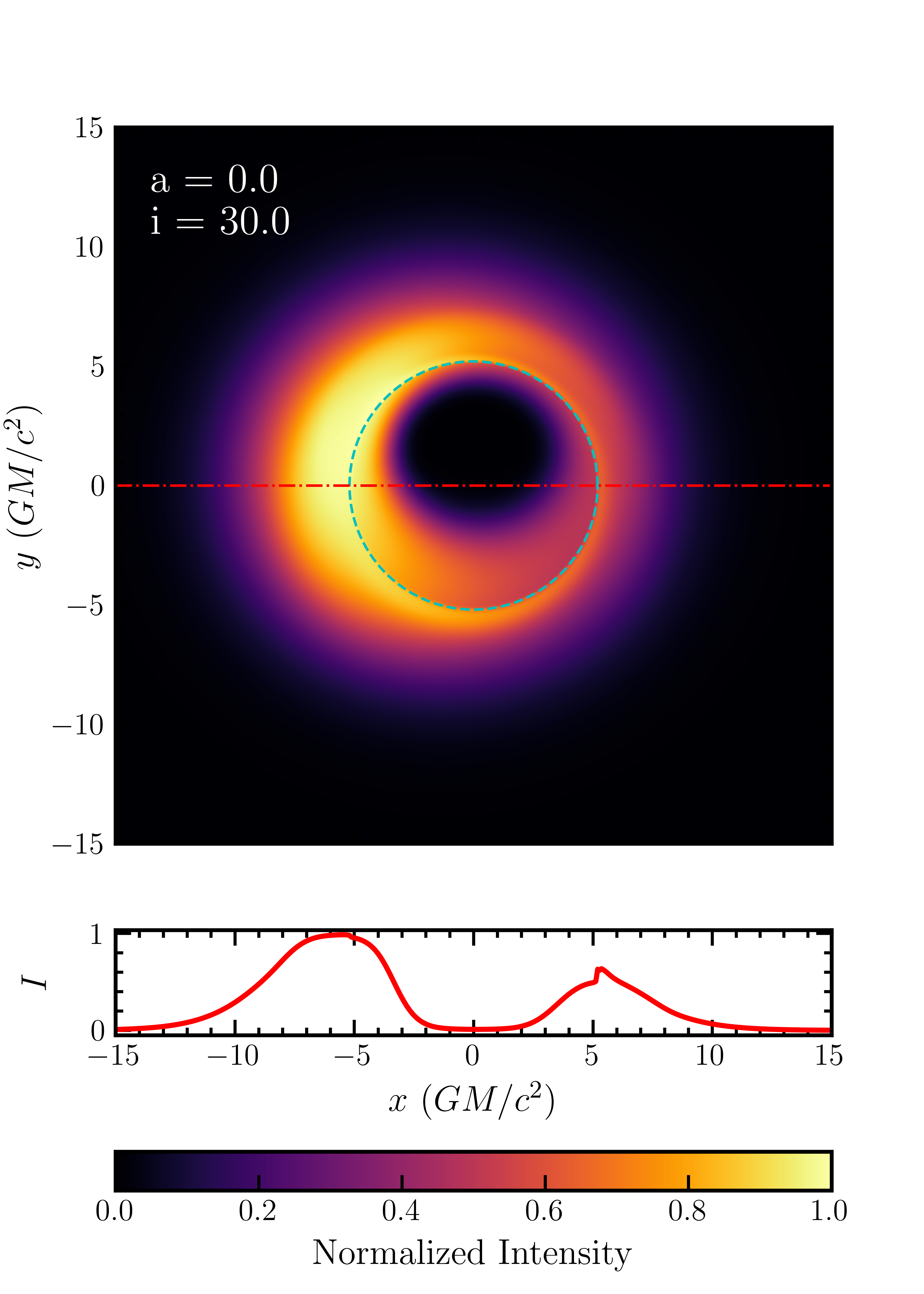}
\includegraphics[width=1.0\columnwidth]{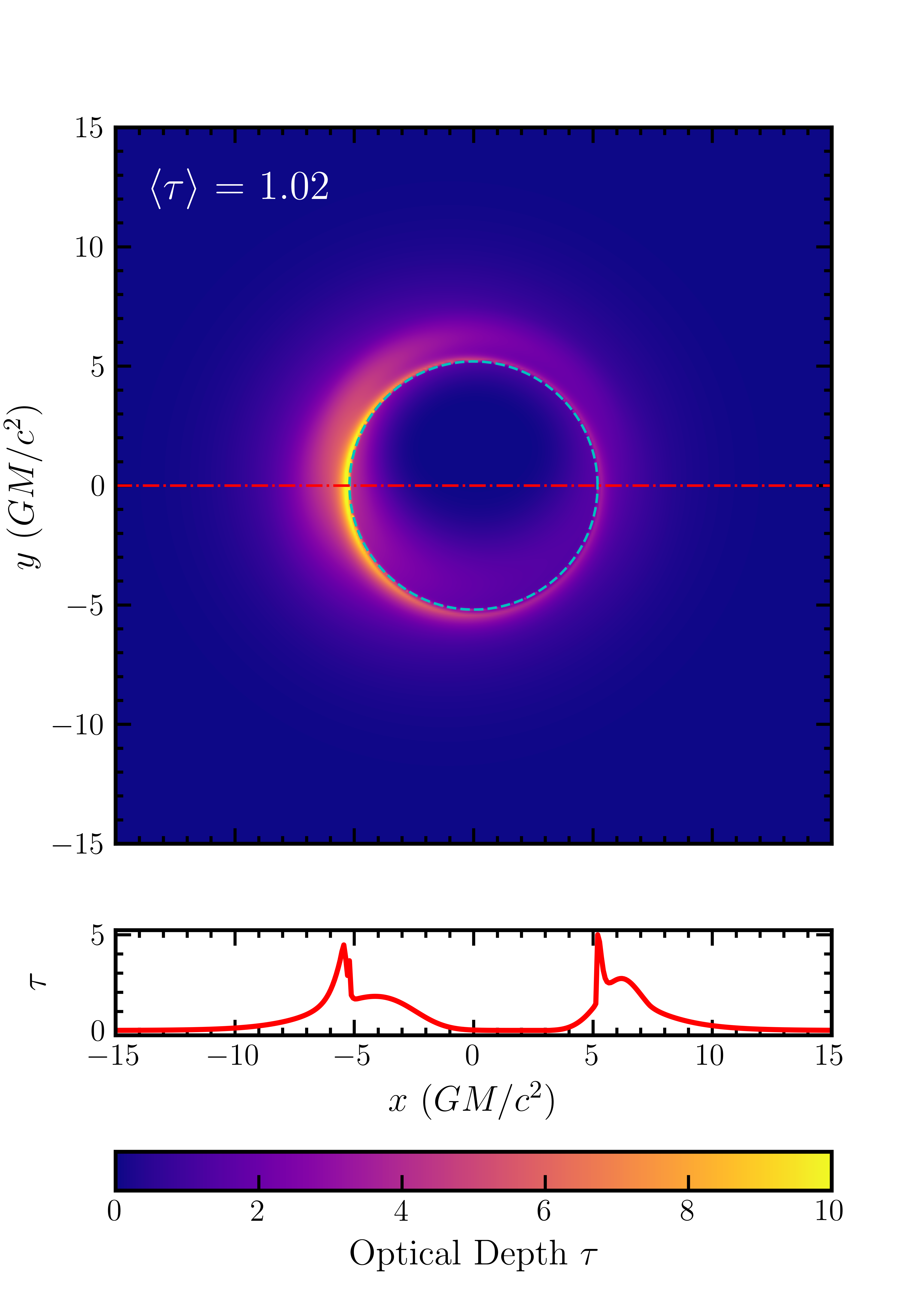}
\caption{{\em (Left)\/} Simulated black hole image and {\em (Right)\/} optical depth along each line-of-sight, along with two horizontal cross sections at $y=0$, shown in the lower panels. Dashed blue curves outline the shadow boundary.}
\label{fig:image_tau}
\end{figure*}

\section{Characterizing the Target Black Holes} 
\label{sec:properties}

We turn to modeling each source using spectral observations and the framework we described in \S\ref{sec:model} to find the transition frequency where the shadow becomes observable. This involves determining the mass accretion rate that reproduces the 1.3~mm flux, calculating the images at a broad range of frequencies,  characterizing their sizes, and identifying the frequency at which the latter becomes comparable to the expected size of the black hole shadow. We do this for every combination of accretion flow and black hole model parameters. We choose to use the flux at 1.3~mm (230 GHz) to normalize the mass accretion rate both because observations at this frequency are uniformly available for targets of interest and because the emission at this frequency is typically dominated by the thermal contribution from the accretion flow rather than from the jet. 

\subsection{Constraining the Mass Accretion Rate}

The mass accretion rate sets the overall density scale of the accretion flow and, therefore, the synchrotron emissivity and opacity. We calculate the spectrum for different accretion rates and identify the rate that reproduces the observed 1.3~mm flux.  

Synchrotron absorption decreases rapidly with increasing frequency. As a result, the total optical depth through a line of sight also decreases rapidly with increasing frequency, causing the corresponding photosphere to move closer to the horizon, where the temperature is also higher. As long as the flow remains optically thick, the resulting brightness, therefore, is limited by the local blackbody function at the local photosphere and continues to increase with increasing frequency. Increasing the mass accretion rate moves the photosphere at each frequency to larger distances from the black hole, thereby increasing the overall flux. 

At some critical frequency, most lines-of-sight become optically thin throughout the accretion flow, enabling the unobstructed observation of the black-hole shadow. At a slightly larger frequency, a balance between higher temperature and the decreasing cross-section for synchrotron emission is reached, where a further increase in frequency causes the overall flux from the source to drop off significantly. 

The top panel of Figure~\ref{fig:example} shows synchrotron spectra for different values of black hole mass and mass accretion rate. We set $M = 1 \times 10^8 M_{\odot}$ and $\dot{M} = 10^{-5} \dot{M}_{\rm Edd}$ as fiducial values and vary one of these parameters at a time. As discussed above, decreasing the accretion rate decreases the overall flux and shifts the optically thick-to-thin transition to lower frequencies. The mass of the black hole has the opposite effect on the optical depth and spectrum. We show in Appendix~\ref{sec: Appendix B} that the transition frequency scales as $\propto \dot{m}^{2/3}/M^{1/3}$, where $\dot{m} \equiv \dot{M}/\dot{M}_{\rm Edd}$. 

For each source and for each combination of the parameters in Table~\ref{tab:parameters}, we use \texttt{KRAY} to perform general relativistic radiative transfer through the accretion flow to obtain an array of the intensities and optical depths at 230~GHz. Using the available measurements for the mass and distance of each target that we compiled in Table~\ref{tab:candidates}, we integrate the intensity over the image plane and adjust the mass accretion rate to match the observed flux. We then use this mass accretion rate to calculate images from 10$-$1000~GHz. 

\begin{figure}[t]
    \centering
    \includegraphics[width=0.93\linewidth]{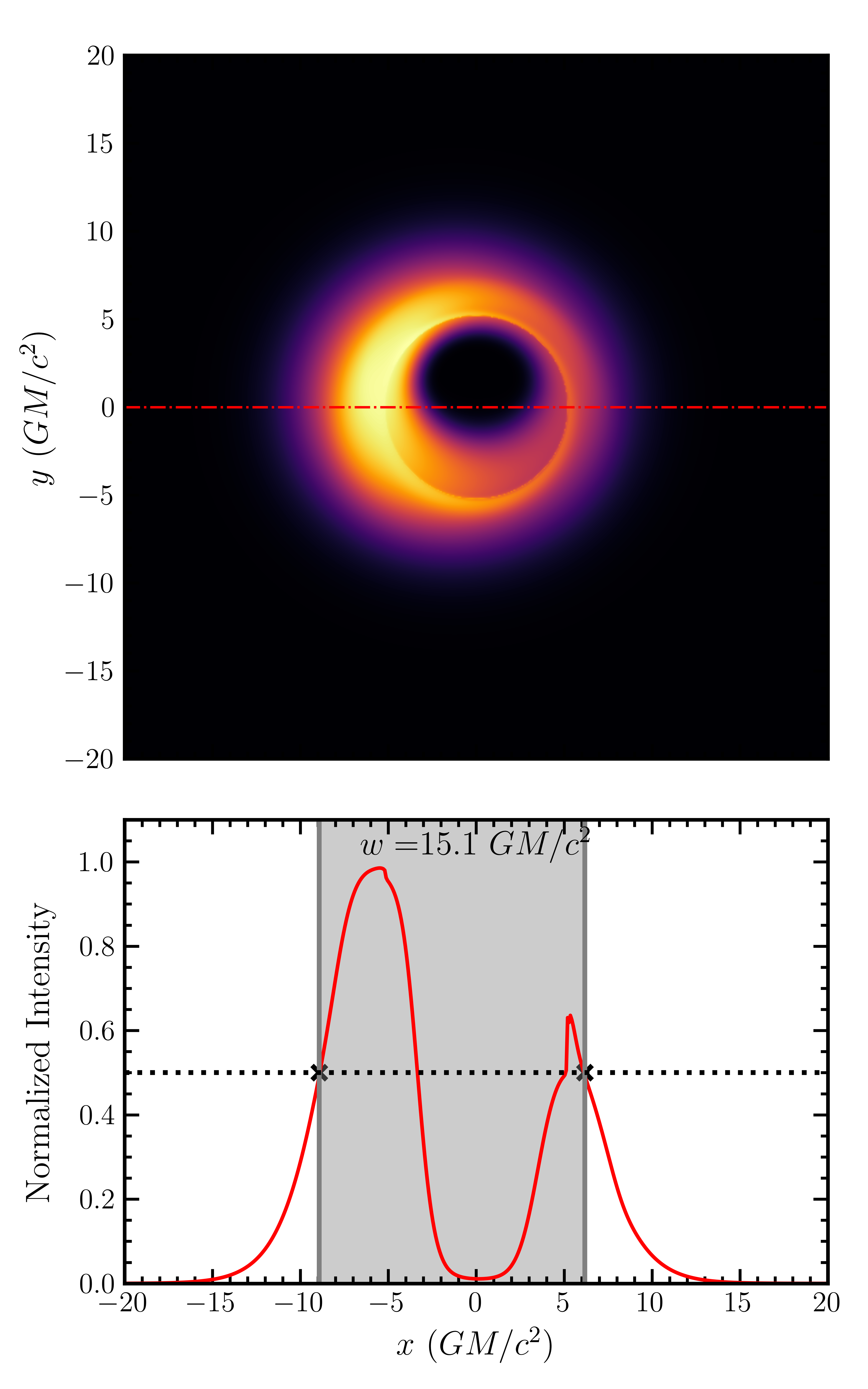}
    \caption{{\em (Top)\/} Example black hole image and (Bottom) its horizontal cross-section at $y=0$. The measurement of the image diameter is indicated by the outer boundaries of the gray-shaded region.}
    \label{fig:imagewidth}
\end{figure}

\subsection{Image Characteristics}

We explore in two ways whether the images we calculate for each combination of model parameters have an unobscured black hole shadow.  

First, we calculate the average optical depth between the observer and the black hole for lines of sight that are within the boundary of the black hole shadow. When this optical depth is high, the shadow is obscured by intervening material. The boundary of the shadow is a function of both the observer inclination and the spin of the black hole. For a large range of black hole spins ($a\lesssim 0.95 $), the shadow remains circular with its center shifted from the line-of-sight to the geometric center \citep{JohansenCenterShift}. We use the fitting formula for the shadow diameter devised by \citet{GRAY} 
\begin{equation}
    r_{\rm sh} \simeq R_0 + R_1 \text{cos}(2.14i -22.9^{\circ})\;,
\end{equation}
where 
\begin{align}
    R_0 =&\,5.2 - 0.209a +0.445a^2 - 0.567a^3\\ 
    R_1 =& \, \left[ 0.24 - \frac{3.3}{(a-0.9017)^2 +0.059}\right] \times 10^{-3}\;.
\end{align}
We then calculate the average optical depth inside the boundary as
\begin{equation} 
\label{eq:averageopticaldepth}
    \langle\tau\rangle =  \frac{1}{N}\sum_{r_{ij}<r_{\text{sh}}} \tau(r_{ij})\;.
\end{equation}
The bottom panel of Figure~\ref{fig:transitionfrequency} show the frequency dependence of the average optical depth, for different values of the black hole mass and accretion rate. As expected, the optical depth drops with increasing frequency, and becomes less than unity at a frequency that is 50\% smaller than the peak frequency of the spectrum. 

\begin{figure}[t]
    \centering
    \includegraphics[width=0.98\columnwidth]{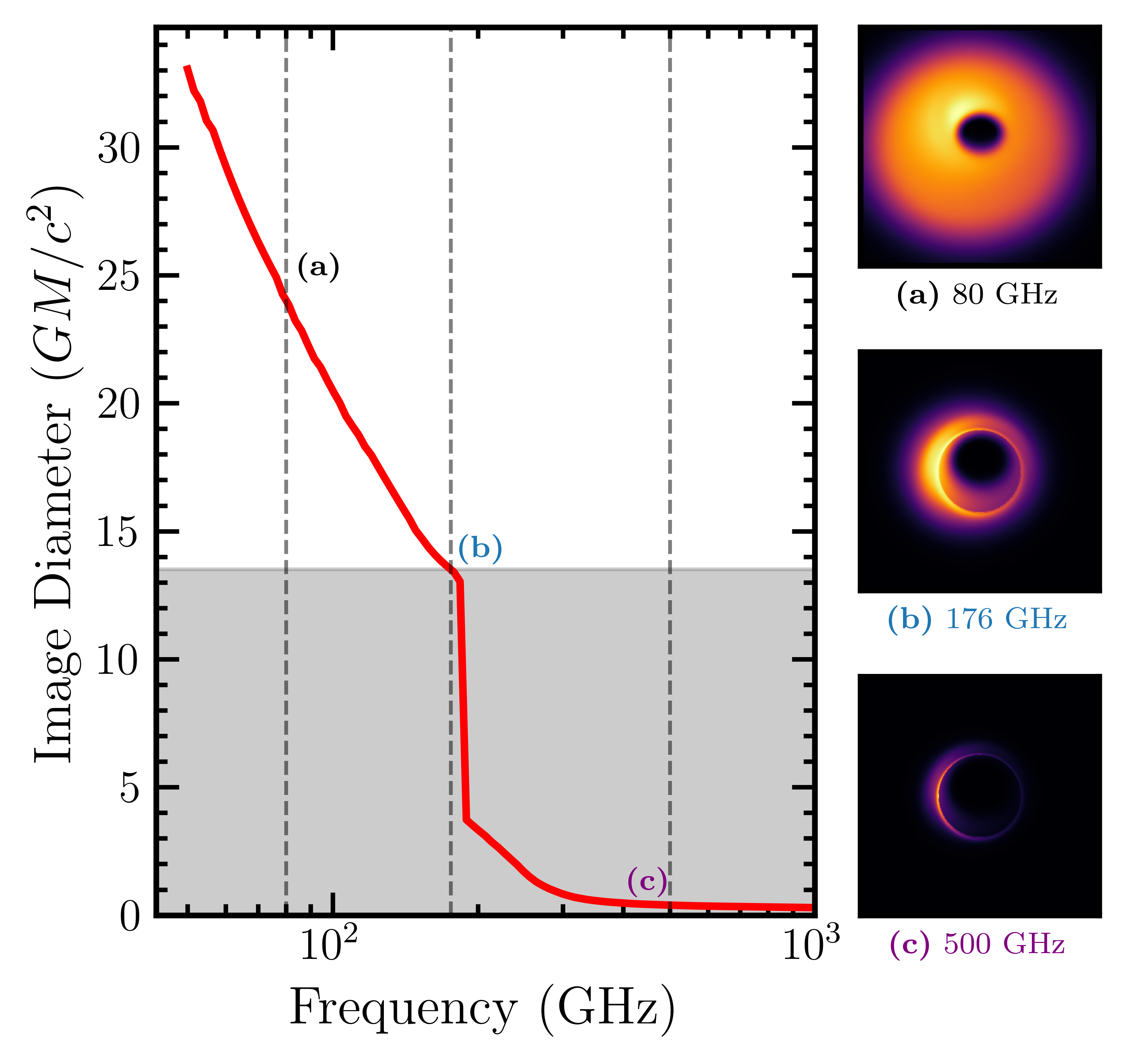}
    \caption{The dependence of image diameter on frequency of observation, for the accretion flow shown in Fig.~\ref{fig:imagewidth}; the right panels show three representative images. The gray-shaded region corresponds to image diameters smaller than 1.3 times the diameter of the black-hole shadow.}
    \label{fig:diameter_vs_f}
\end{figure}

We show in Figure~\ref{fig:image_tau} an example simulated 230~GHz black hole image and the corresponding optical depth. In the subpanels, we also show the horizontal cross sections of the image brightness and optical depth. For this configuration, the average optical depth is $<\tau>=1.02$. The cross section of the optical depth shows that this average may become dominated by the peaks near the shadow boundary, as one would expect from the rapidly increasing pathlength at those impact parameters. For this reason, the average optical depth may not be the most accurate indicator of the visibility of the shadow in a black hole image.  

Because of this, we introduce a second method to quantify obscuration. We characterize an image by calculating its frequency-dependent diameter along the cross-section perpendicular to the black hole spin. Using this cross-section, we find the maximum intensity and calculate the FWHM of the image, taking the outermost points where the intensity is equal to half of the maximum value (see Figure~\ref{fig:imagewidth}). We expect the image width to decrease as the observing frequency increases and the $\tau =1$ surface moves toward the black hole. 

\begin{figure*}[!ht!]
    \centering
    \includegraphics[width=0.95\linewidth]{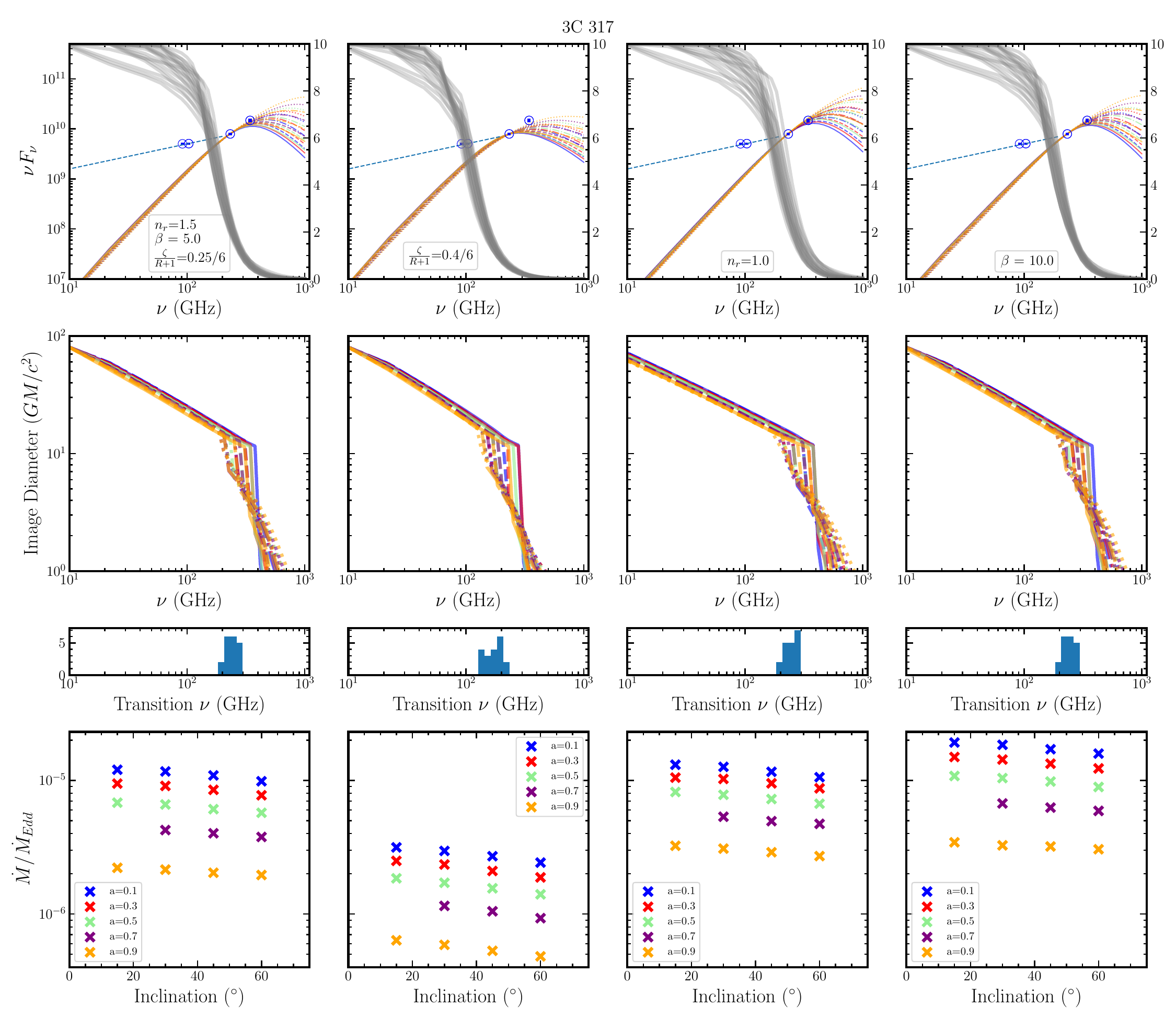}
    \caption{{\em (Top row)\/} Illustrative observations (data points), spectral models (orange), and corresponding average optical depths (gray) for 3C~317 and different black-hole spins and observer inclinations. Different combinations of accretion flow parameters (electron temperature,  $\zeta/(R+1)$, radial velocity profile power law, $n_r$, and plasma $\beta$) are shown in each column. Blue dashed lines show the estimated jet contribution to the flux at low frequencies. {\em (Second row)\/} Image diameter as a function of observing frequency. {\em (Third row)\/} Distribution of transition frequencies beyond which an unobstructed black-hole shadow can be observed. {\em (Bottom row)\/} Inferred mass accretion rate in units of $\dot{M}_{\rm Edd}$ as a function of observer's inclination. }
    \label{fig:ex3C317}
\end{figure*}

Figure~\ref{fig:diameter_vs_f} shows the dependence of the image diameter on frequency. The image diameter tracks the outer boundary of the photosphere. At low frequencies, the $\tau =1$ surface is far out in the accretion flow, corresponding to a large image. As the frequency increases, the effects of self-absorption decrease as does the diameter of the image. It eventually reaches a limit imposed by the physical size of the shadow. At even higher frequencies, Doppler boosts (at high observer inclinations) cause one side of the image to be significantly brighter than the other. In this case, our definition of the image diameter measures predominantly the size of the Doppler-boosted region of the image and is smaller than the shadow diameter. We use image diameter to define the critical frequency as the one at which the inferred image diameter becomes 30\% larger that of the black-hole shadow.

\section{Optimal Observing Frequency for Shadow Imaging} 
\label{sec:spectra}

We analyze the sources in Table~\ref{tab:candidates} using the methodology we discussed in the previous sections. In terms of observability down to the horizon scales, we find that the sources can be grouped into two categories. The first group has a transition frequency that lies in the $200-500$~GHz for all combinations of model parameters. There are some differences between sources among this group: for one subcategory, the shadow can consistently be imaged at lower (200$-$300~GHz) frequencies, while for a second subcategory there is a stronger dependence on model parameters; this subcategory may require observations in the $400-500$~GHz range to image horizon scales. The second category contains sources for which the emission is optically thick up to 1000~GHz or the spectrum cannot be modeled with a radiatively inefficient accretion flow model. This could happen if there is substantial contribution to the image core at 230~GHz by the jet or the mass accretion rate is too high for the flow to be radiatively inefficient.

\begin{figure*}[!ht!]
    \centering
    \includegraphics[width=0.9\linewidth]{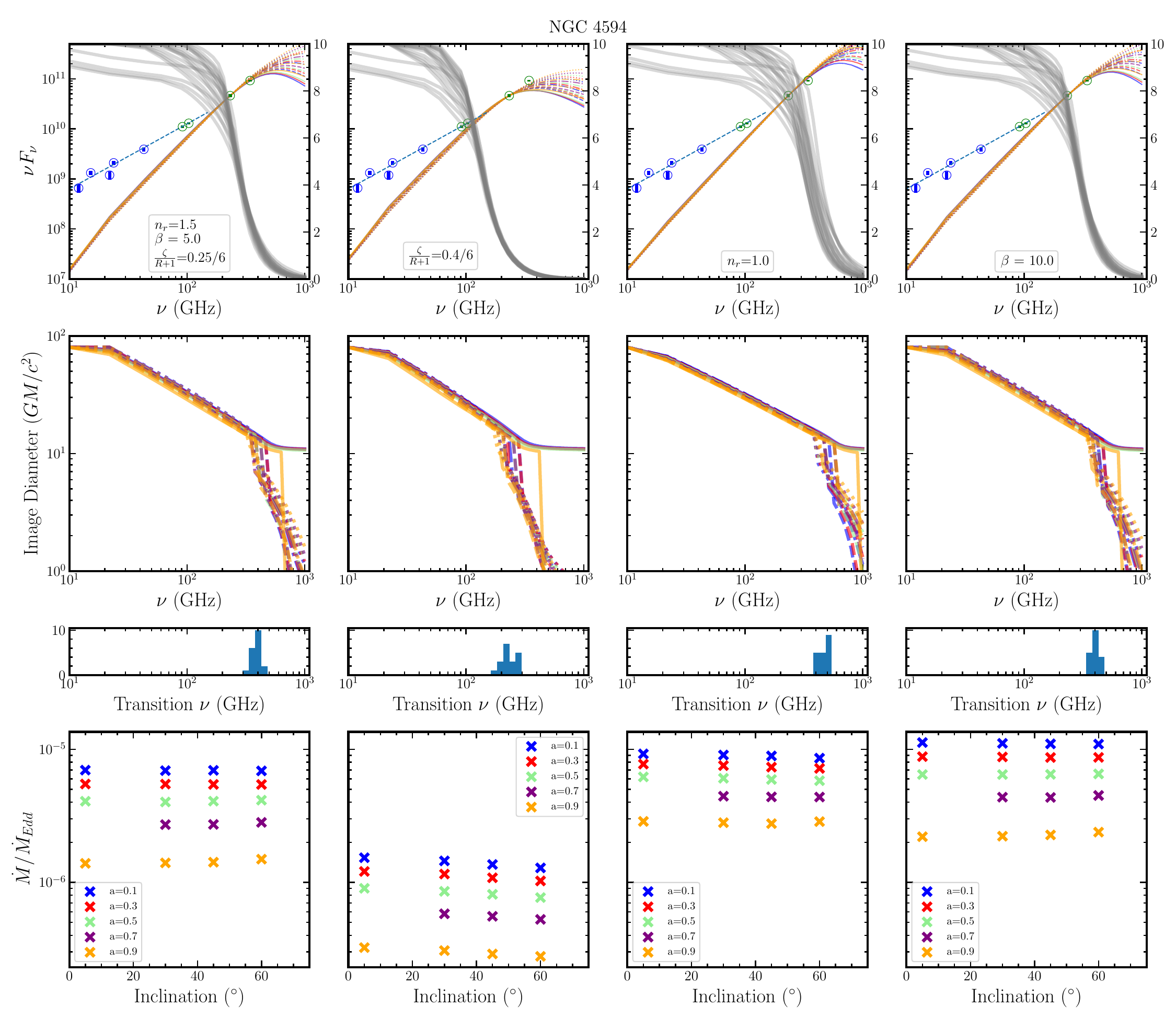}
    \caption{Same as Fig.\ref{fig:ex3C317} but for NGC~4594.}
    \label{fig:exNGC4594}
\end{figure*}

\subsection{Low Transition-Frequency Sources}

For the majority of the supermassive black holes targets we analyzed, the transition frequency is in the 200$-$500~GHz for all combinations of model parameters. These include: 1C~1459, NGC~2663, NGC~3894, NGC~4552 in the first subcategory and NGC~4594, NGC~4261, M~84, and NGC~1399 in the other. This practically guarantees that their shadows would be observable at these frequencies as long as the interferometer has sufficient angular resolution and sensitivity. 

As an example in the first subcategory, we show here in detail the analysis for 3C~317, a source with a well-studied spectrum across a broad range of radio frequencies. We discuss the remaining sources in Appendix~\ref{sec:Appendix A}. Figure~\ref{fig:ex3C317} shows spectral models for 3C~317 for different spins and inclinations and four illustrative combinations of the parameters describing the electron temperature (second column), radial velocity profile (third column), and plasma $\beta$ (fourth column). The characteristic transition frequency, inferred from the sudden drop in the optical depth (top row) or the image diameter (second row) is insensitive to the black hole and model parameters. 

The ratio of electron-to-ion temperature has the most significant effect on the transition frequency. The higher electron temperatures necessitate a smaller size of the photosphere and hence a lower mass accretion rate to reproduce the observed 230~GHz flux, thereby also reducing the transition frequency. For the radial profiles, a higher value of the power-law index $n_r$ increases the total optical depth for the same accretion rate, moving the transition frequency to somewhat higher values. Finally, the transition frequency is practically insensitive to the plasma parameter $\beta$. 

As an example in the second subcategory, we show in Figure~\ref{fig:exNGC4594} the spectrum and models for NGC~4594, also known as M104 or the Sombrero Galaxy, which is a relatively nearby spiral that has been well-observed in the radio. The range of transition frequencies for this source is larger, depending on model parameters, possibly requiring observations above what has traditionally been carried out for VLBI imaging. 

\subsection{Optically-Thick Sources}

\begin{figure*}[!ht!]
    \centering
    \includegraphics[width=0.9\linewidth]{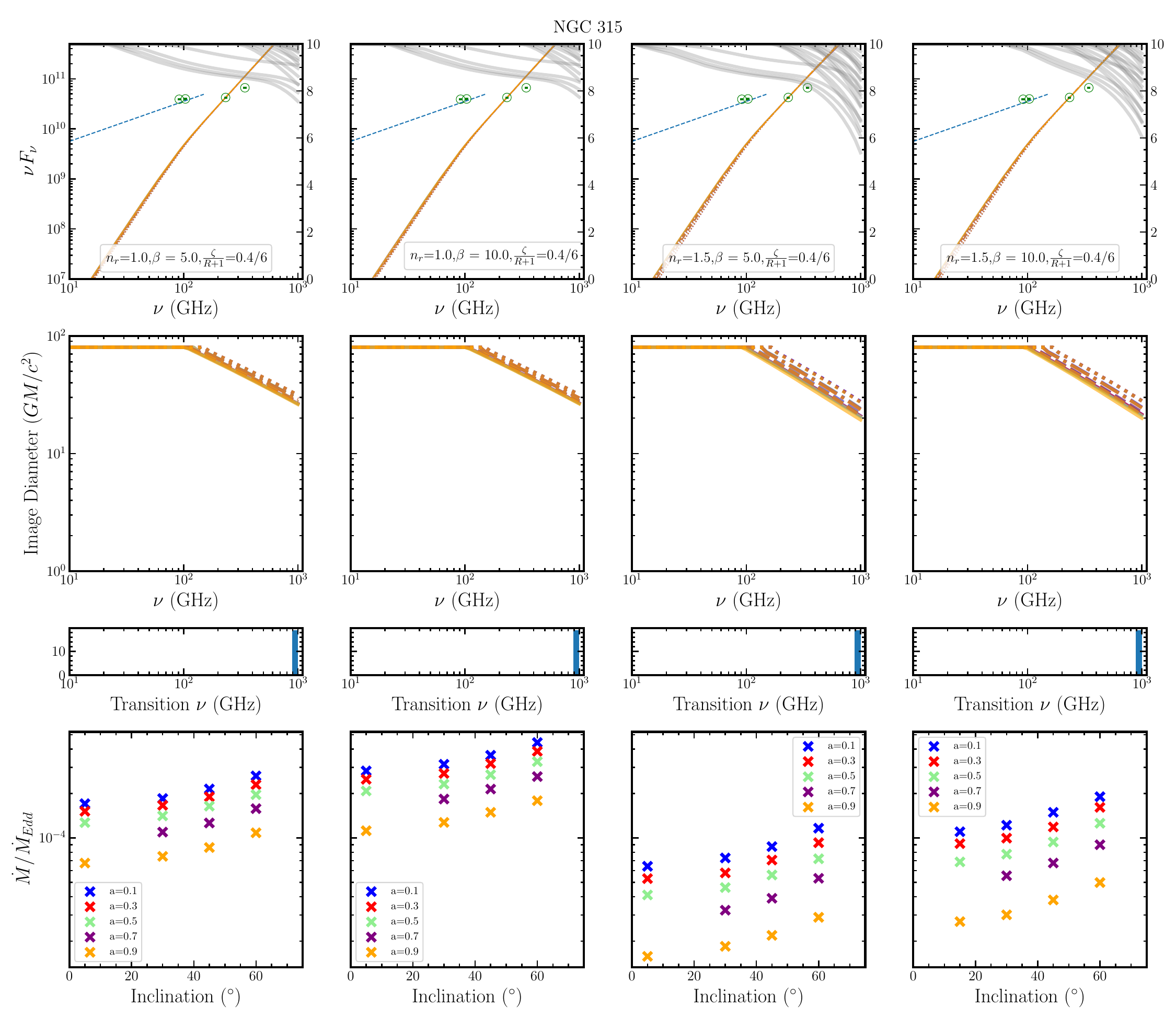}
    \caption{Same as Fig.\ref{fig:ex3C317} but for NGC~315. Only models with high electron temperatures $\zeta/(R+1) =0.4/6$ provide meaningful fits.}
    \label{fig:exNGC315}
\end{figure*}

We found the emission from several black holes in our target list to be optically thick for the entire range of frequencies we studied. As an example, we show in Figure~\ref{fig:exNGC315} the spectrum and models for NGC~315. 

The large optical depths we infer for these sources may result from one of three causes. First, these sources may have mass accretion rates that are higher than those of the previous category, causing the flows to be optically thick over a larger range of frequencies. Second, the accretion rate might be so high that the flows are, in fact, not radiatively inefficient and may need to be described by a geometrically thin, optically thick accretion disk model.

The third possibility is that the jet component of the source emission, even at 230 GHz, is significant with respect to that of the accretion flow. Even at the best resolutions obtained with existing radio interferometer arrays, such as ALMA and the Submillimeter Array (SMA), the observations cannot fully  separate the non-negligible flux contributions originating from the jet component from that of the accretion flow. Because we model only the flux originating from the accretion flow, attempting to fit the total unresolved flux would result in overcompensating with a higher mass accretion rate. 

In any of these cases, high-resolution images of the energetic jets as well as image size measurements can still provide useful insight into accretion processes in this regime of accretion rate.

\begin{figure*}[ht]
    \centering
    \includegraphics[width=0.96\linewidth]{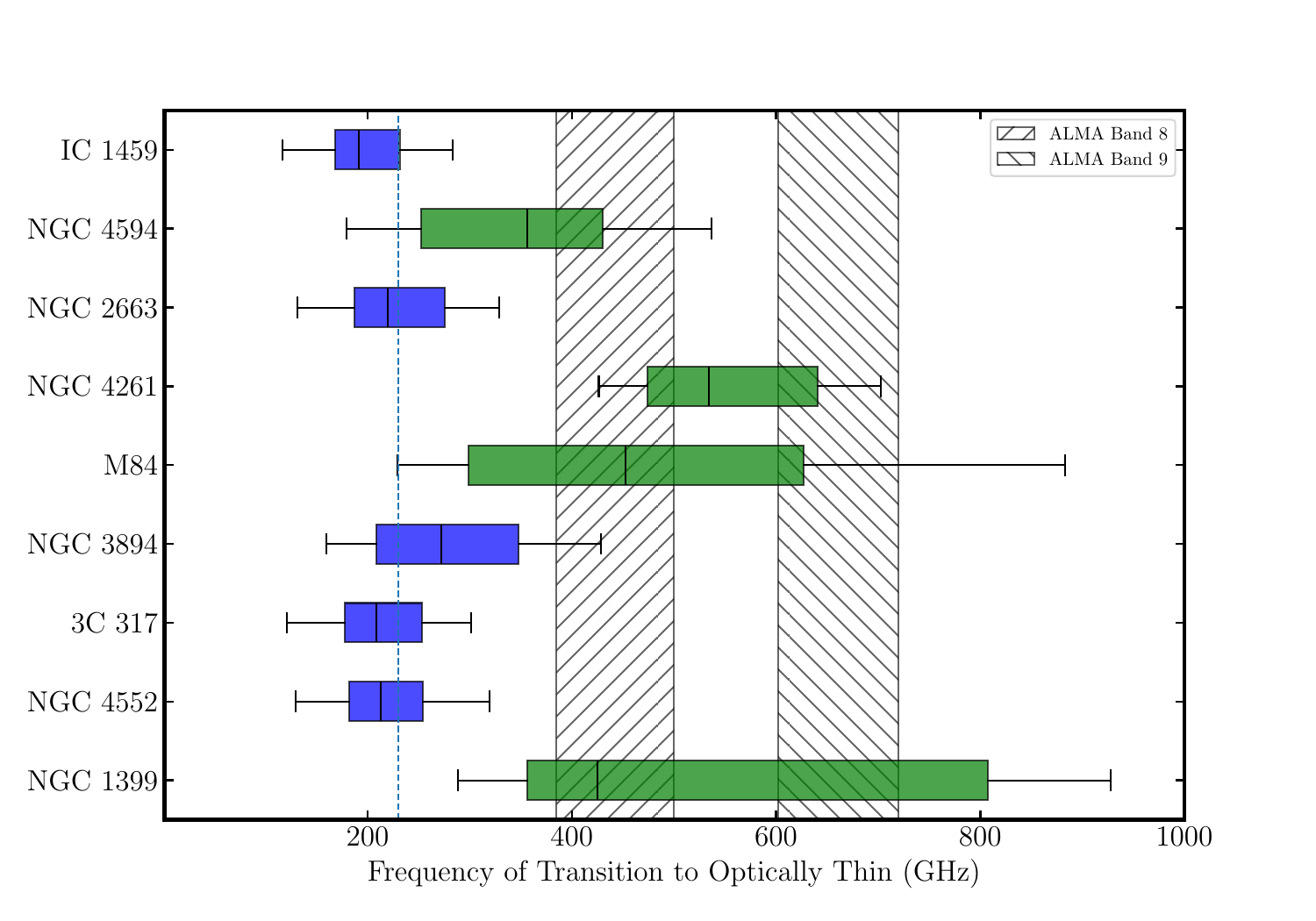}
    \caption{Error bars show the ranges of the transition frequency from optically thick to optically thin emission for the combination of model parameters described in Table~\ref{tab:parameters}. Colored bars for each source range from the 25-th to the 75-th percentiles of the distribution of transition frequencies. The dotted blue line shows the 230~GHz frequency of the initial EHT observations. High-frequency ALMA Bands 8 \& 9 are shown in hatched regions for reference to ground-based radio telescopes. Blue colors correspond to sources with 200-300~GHz transition frequencies, while green colors correspond to sources with larger transition frequencies.}
    \label{fig:transitionfrequency}
\end{figure*}

\section{Conclusions}\label{sec:conclusions}

We evaluated supermassive black-hole targets for horizon-scale imaging based on their compact radio fluxes at 230 GHz, their masses, and their distances. Starting with a list of bright nearby sources with large angular diameters, we modeled the emission of their accretion flows using a flexible, fully covariant semi-analytic model and general relativistic radiative transfer. We explored a wide variety of properties for both the accretion flows and the black holes and characterized the images to determine prospects for imaging their shadows. 

Figure~\ref{fig:transitionfrequency} shows the range of frequencies at which the accretion flows become optically thin for each source, for the various combinations of model parameters we explored here. Observations at these frequencies would allow for imaging of their unobscured black-hole shadows. This figure does not include the four sources for which the models did not result in optically thin accretion flows in this range of frequencies. 

The two subcategories of sources are evident in this figure. For IC~1459, NGC~2663, NGC~3894, 3C~317, and NGC~4552, the sensitivity to model, black hole, and observer parameters is weak, resulting in a small range of transition frequencies that are in the 200$-$300~GHz range. The second subgroup includes NGC~4594, NGC~4261, M84, and NGC~1399, for which there is a broader range of potential transition frequencies, and require higher observing frequencies for shadow imaging. For example, observations at frequencies comparable to ALMA Band 8 would meet this requirement for these sources.

This modeling and characterization can help guide the choice of observing frequencies for advanced radio interferometry missions from space (see, e.g., \citealt{EHimager,Millimetron,Capella,BHEX_Motivation,ChineseSpaceMillimeterVLBI,zineb2024advancing}), for which the wavelength will impact not only the angular resolution and the sensitivity of the array, but also the observability of the targets. 

\appendix

 \section{Observations of Black Hole Targets}
\label{sec:Appendix A}

In this Appendix, we provide a more detailed list of source properties and observations for each source we analyzed in this paper. We focus, in particular, on the existing flux measurements in the radio band as well as the masses and distances of the supermassive black holes. We provide errors for the flux measurements when available. 

In some cases, we identify whether the emission originates from the central or core region. When not stated, we assumed that the low-frequency emission is not resolved, whereas the high-frequency measurements are dominated by the core. 

\begin{figure}
    \centering
    \includegraphics[width=0.99\linewidth]{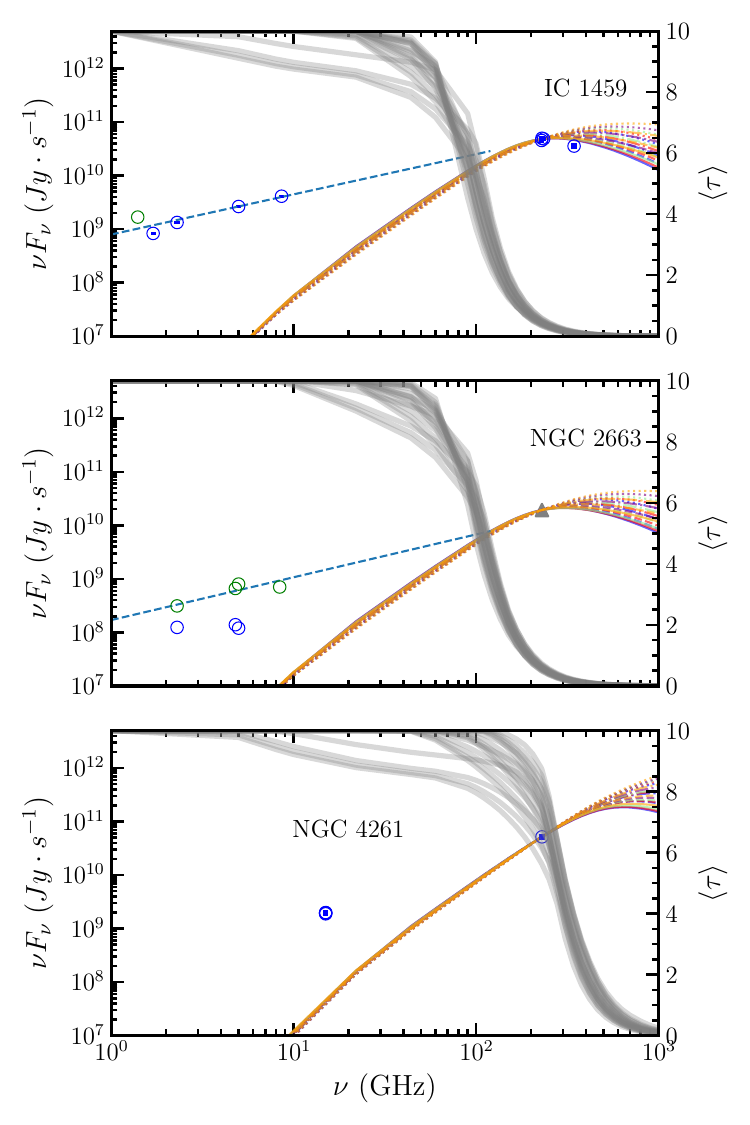}
    \caption{Observed fluxes (points) and  representative models (orange curves) for the supermassive black holes IC 1459, NGC 2663, and NGC 4261. Gray lines show the corresponding average optical depths $\langle\tau\rangle$ as a function of frequency. Data that correspond to core emission are shown in blue, whereas data for unresolved emission is shown in green. Blue dashed lines show the estimated jet contribution to the flux at low frequencies.\label{fig:spectraset1}} 
\end{figure}

\subsection{IC 1459}

IC 1459 is an elliptical galaxy that is 26.2~Mpc away \citep{CosmicFlows2}, with a strong counter-rotating stellar component associated with the core of the galaxy \citep{IC1459-2}. The mass of the central supermassive black hole, measured from stellar kinematics, is $2.6 \pm 1.1\times10^9 \, M_{\odot}$ \citep{IC1459mass}. The core of the galaxy hosts a radio source with near symmetric jets \citep{IC1459Flux2}. Table~\ref{tab:IC1459} shows the compiled observations across radio frequencies. 

\begin{deluxetable}{CCCCC} [t!]
    \tablecaption{Radio Observations of IC1459
    \label{tab:IC1459}}
    \tablehead{
        \colhead{$\nu$ (GHz)} & \colhead{Flux (mJy)} & \colhead{Telescope} & \colhead{Region} & \colhead{Reference}
    }
    \startdata
    1.4   & 1200   & \text{VLA}  & \text{Total} & 1 \\
    1.7   & 490  \pm 30        &   \text{VLBA} &  \text{Core} & 2 \\
    2.3   & 580  \pm 30        &  \text{VLBA} &  \text{Core} &  2\\
    5.0   & 530  \pm 30        &  \text{VLBA} &  \text{Core} & 2 \\
    8.6   & 480  \pm 30        &  \text{VLBA} &  \text{Core} &  2\\
    228.9 & 198.4 \pm 20.3     &  \text{SMA} &  \text{Core}  & 3 \\
    230.0 & 217.0 \pm 21.0     &  \text{ALMA} &  \text{Total} &  1\\
    234.9 & 206.7 \pm 21.2     &   \text{SMA}&  \text{Core} & 3 \\
    344.9 & 103.4 \pm 13.7     &   \text{SMA} &   \text{Core} & 3 \\
    \enddata
\tablerefs{1.\ \citet{IC1459Flux1}; 2.\ \citet{IC1459Flux2}; 3.\ \citet{SMAFluxSurvey} }
\end{deluxetable}

\subsection{NGC 4594}

NGC 4594, also known as M104 or the Sombrero Galaxy, is a well-studied spiral galaxy in the Southern Sky. The galaxy is located $9.55\pm 0.13 \pm 0.31$~Mpc away from Earth (\citealt{M104Distance}), which was measured using the tip of the red-giant branch (TRGB) method. The central black hole has a mass of $6.6 \pm 0.4\times10^8 \, M_{\odot}$ measured from globular cluster kinematics (\citealt{M104Mass}). Table~\ref{tab:NGC4594} compiles observations across the radio spectrum. The supermassive black hole has an associated relativistic jet. Different studies have reported different estimates of the jet inclination: \citet{M104Jet1} found $\leq 25^{\circ}$ whereas \citet{M104Jet2} found $66^{\circ \, +4^{\circ}}_{\; \; \, -6^{\circ}}$. 

\begin{deluxetable}{CCCCC}[t]
    \tablecaption{Radio Observations of NGC4594\label{tab:NGC4594}}
    \tablehead{
        \colhead{$\nu$ (GHz)} & \colhead{Flux (mJy)} & \colhead{Telescope} & \colhead{Region} & \colhead{Reference}
    }
    \startdata
    1.4   & 59.6 \pm 6.7    & \text{VLBA} & \text{Core}  & 1 \\
    2.3   & 62.1 \pm 6.2    & \text{VLBA} & \text{Core}  & 1 \\
    5.0   & 74.3 \pm 7.4    & \text{VLBA} & \text{Core}  & 1 \\
    8.4   & 80.2 \pm 8.0    & \text{VLBA} & \text{Core}  & 1 \\
    12.0  & 54.7 \pm 10.9   & \text{VLBA} & \text{Core}  &  2\\
    15.2  & 87.1 \pm 8.7    & \text{VLBA} & \text{Core}  &  1\\
    22.0  & 54.1 \pm 10.8   & \text{VLBA} & \text{Core}  &  2\\
    23.8  & 88.0 \pm 8.8    & \text{VLBA}& \text{Core}  &  1\\
    43.0  & 91.0 \pm 9.1    & \text{VLBA} & \text{Core}  &  1\\
    91.5  & 121.0 \pm 6.0   & \text{ALMA} & \text{Total} & 3 \\
    103.5 & 125.0 \pm 6.0   & \text{ALMA} & \text{Total} & 3 \\
    233.0 & 198.0 \pm 13.0  &  \text{ALMA}& \text{Total} & 3 \\
    343.5 & 269.0 \pm 10.0  & \text{ALMA} & \text{Total} & 3 \\
    \enddata
\tablerefs{1.\ \citet{M104Jet1}; 2.\ \citet{M104Jet2}; 3.\ ALMA Calibrator Catalog}
\end{deluxetable}

\subsection{NGC 3998}

NGC~3998 is a nearby LINER (low ionization nuclear emission region) galaxy with an associated, low-power radio active galactic nucleus \citep{NGC3998Flux1}. It has radio lobe structures emanating from the central supermassive black hole and a kpc-scale jet on the northern side of the core \citep{NGC3998Flux4}. The central mass of the black hole is $8.1^{+2.0}_{-1.9} \times 10^8 \;M_{\odot}$ determined from stellar dynamics \citep{NGC3998Mass}. The galaxy is located $14.2\,$ Mpc away \citep{CosmicFlows2}, determined from Surface Brightness Fluctuation (SBF) methods. Table~\ref{tab:NGC3998} compiles observations across the radio spectrum. Because we do not have a prior flux measurement at 230~GHz, we use the one measured at 43~GHz and extrapolate to 230~GHz assuming a flat spectrum.

\begin{deluxetable}{CCCCC}[h]
    \tablecaption{Radio Observations of NGC3998\label{tab:NGC3998}}
    \tablehead{
        \colhead{$\nu$ (GHz)} & \colhead{Flux (mJy)} & \colhead{\text{Telescope}} & \colhead{\text{Region}} & \colhead{\text{Reference}}
    }
    \startdata
    1.4   & 109 \pm 3    & \text{VLA} & \text{Total}  & \text{3} \\
    1.4   & 175 \pm 9 & \text{WSRT*} & \text{Core}  & \text{4} \\
    1.4   & 177 \pm 9    & \text{WSRT} & \text{Core}  & \text{4} \\
    1.4   & 149 \pm 7    & \text{WSRT} & \text{Core}  & \text{4} \\
    1.5   & 107 \pm 3    & \text{VLA} & \text{Total}  & \text{1} \\
    1.5   & 98  \pm 3    & \text{VLA} & \text{Total}  & \text{1} \\
    4.9   & 98  \pm 3    & \text{VLA} & \text{Total}  & \text{1} \\
    4.9   & 83  \pm 3    & \text{VLA} & \text{Total}  & \text{1} \\
    4.9   & 88  \pm 8    & \text{GBT*} & \text{Total}  & \text{2} \\
    5.0   & 118 \pm 6    & \text{WSRT} & \text{Total}  & \text{4} \\
    43.0  & 133         & \text{VLBA} & \text{Total} & \text{5} \\ 
    \enddata
    \tablerefs{ 1.\ \citet{NGC3998Flux1}; 2.\ \citet{NGC3998Flux2}; 3.\ \citet{NGC3998Flux3}; 4.\ \citet{NGC3998Flux4}; 5 \citet{ngEHTsource+modelling}}
    \tablecomments{* Westerbork Synthesis Radio Telescope (WSRT), **Green Bank Telescope (GBT)}
\end{deluxetable}

\subsection{NGC 2663}

NGC 2663 is an elliptical galaxy that is located 28.5~Mpc away, determined using the TFR method \citep{NGC2663Distance}. It has highly collimated jets, the emission from which extends 355~kpc across \citep{NGC2663Jet}. The black hole mass was estimated to be $1.6 \times 10^9 \;M_{\odot}$ in \citet{ngEHTsource+modelling}, using the measured velocity dispersion from \citet{sigma2011} and the $M$-$\sigma$ relationship from \citet{msigma2016}. Table~\ref{tab:NGC2663} lists observations across the radio spectrum. Due to the sparse observations in the radio, we extrapolate the flux measured at 8.4~GHz to 230~GHz, assuming a flat spectrum. 

\begin{deluxetable}{CCCCC}[h]
    \tablecaption{Radio Observations of NGC 2663\label{tab:NGC2663}}
    \tablehead{
        \colhead{$\nu$ (GHz)} & \colhead{Flux (mJy)} & \colhead{\text{Telescope}} & \colhead{\text{Region}} & \colhead{\text{Reference}}
    }
    \startdata
    2.3   & 136         & \text{VLBA} & \text{Total}  & \text{1} \\
    2.3   & 54          & \text{VLBA } & \text{Core}  & \text{1} \\
    4.8   & 138         & \text{VLBA} & \text{Total}  & \text{1} \\
    4.8   & 29          & \text{VLBA} & \text{Core}  & \text{1} \\
    5.0   & 160         & \text{VLA} & \text{Total}  & \text{2} \\
    5.0   & 24          & \text{VLA} & \text{Core}  & \text{2} \\
    8.4   & 84          & \text{VLBA} & \text{Total}  & \text{1} \\
    \enddata
    \tablerefs{1.\ VLBA Calibrator Catalog 2.\ \citet{NGC2663Flux1}}
\end{deluxetable}

\subsection{NGC 4261}

NGC~4261 is an elliptical galaxy located $32.4$~Mpc away from Earth, as measured using the SBF method \citep{CosmicFlows2}. The mass of the central black hole was estimated to be $1.67^{+0.48}_{-0.34} \times 10^9 \; M_{\odot}$ \citep{NGC315NGC4261MassandDistance} from CO gas dynamics. Early studies of jets with the VLBA inferred a viewing angle of $63^{\circ} \pm 3^{\circ}$. A recent investigation by \citet{NGC4261Flux1} with the VLBA found jet inclinations in the range $54^{\circ}-84^{\circ}$. We show in Table~\ref{tab:NGC4261} existing observations in the radio bands.

\begin{deluxetable}{CCCCC}[h]
    \tablecaption{Radio Observations of NGC 4261\label{tab:NGC4261}}
    \tablehead{
        \colhead{$\nu$ (GHz)} & \colhead{Flux (mJy)} & \colhead{\text{Telescope}} & \colhead{\text{Region}} & \colhead{\text{Reference}}
    }
    \startdata
    15    & 126 \pm 12  & \text{VLBA} & \text{Core}  & 1 \\
    15    & 131 \pm 13  & \text{VLBA} & \text{Core}  & 1\\
    15    & 133 \pm 13  & \text{VLBA} & \text{Core}  & 1\\
    15   & 130 \pm 11  & \text{VLBA} & \text{Core}  &  1\\
    230   & 225 \pm 25  & \text{ALMA} & \text{Core}  & 2 \\
    \enddata
    \tablerefs{1.\ \citet{NGC4261Flux1}; 2.\ \citet{NGC315NGC4261MassandDistance}}
\end{deluxetable}

\subsection{M84}

M84 is an elliptical galaxy located 17~Mpc away from the Earth, as measured using the TFR method \citep{CosmicFlows2}, with a low-luminosity active galactic nucleus. The mass of the central supermassive black hole is measured from gas kinematics to be equal to $8.5^{+0.9}_{-0.8} \times 10^8 M_\odot$ \citep{M84Mass}. Several studies have analyzed the jet of M84, which shows two-sided structures \citep{M84Jet1}. Recent studies using data from the VLBA and the East Asian VLBI Network (EAVN) constrained the jet inclination to be $58^{\circ}\,^{+17}_{-18}$ \citep{M84Flux1}. We show in Table~\ref{tab:M84} existing observations in the radio bands.

\begin{figure} \label{fig:spectraset2}.
    \centering
    \includegraphics[width=0.99\linewidth]{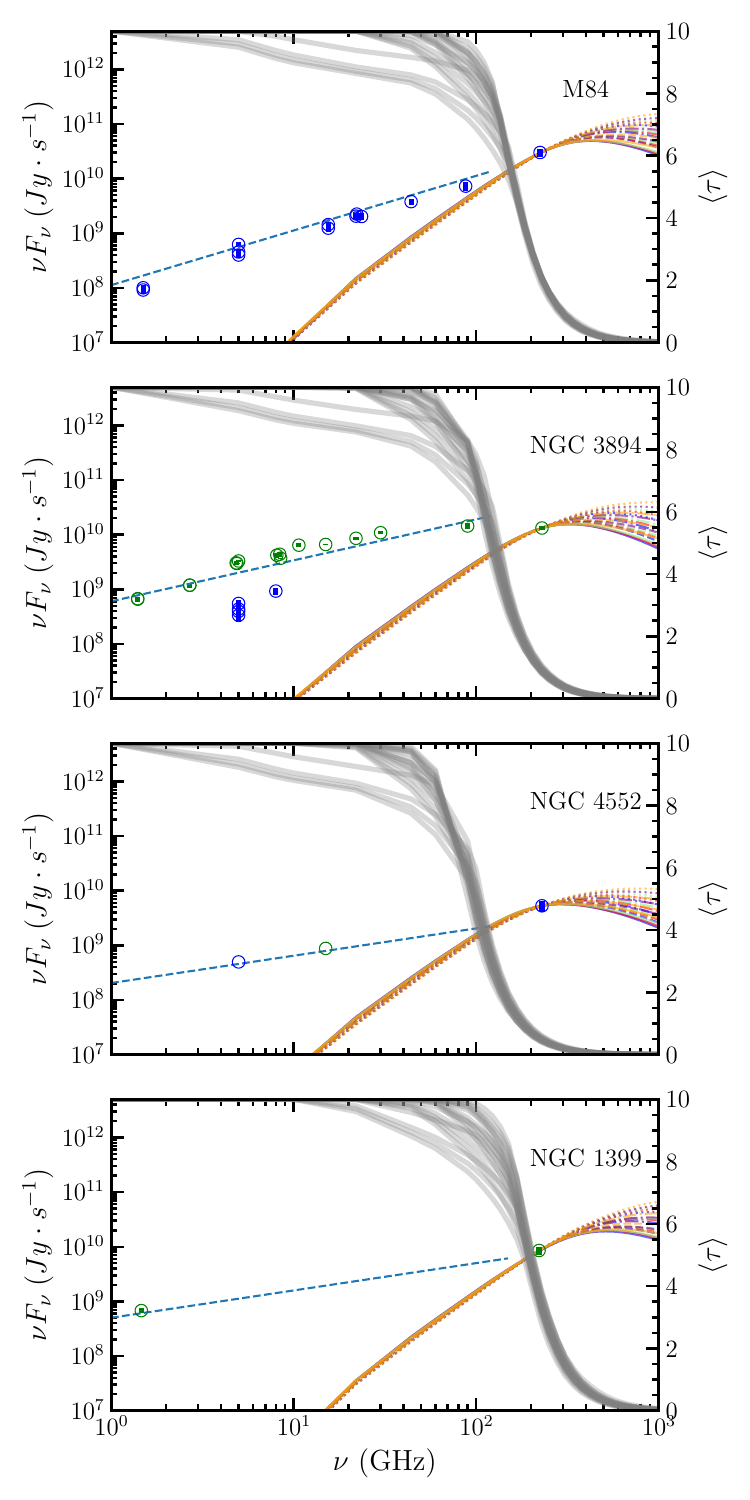}
    \caption{Same as Figure \ref{fig:spectraset1} but for M84, NGC~3894, NGC~4552, and NGC~1399. }
    
\end{figure}
\begin{deluxetable}{CCCCC}[h]
    \tablecaption{Radio Observations of M84\label{tab:M84}}
    \tablehead{
        \colhead{$\nu$ (GHz)} & \colhead{Flux (mJy)} & \colhead{\text{Telescope}} & \colhead{\text{Region}} & \colhead{\text{Reference}}
    }
    \startdata
    1.5   & 61 \pm 9     & \text{VLBA} & \text{Core}  & \text{1} \\
    1.5   & 67 \pm 8     & \text{VLBA} & \text{Core}  & \text{1} \\
    5.0   & 80 \pm 9     & \text{VLBA} & \text{Core}  & \text{1} \\
    5.0   & 91 \pm 10    & \text{VLBA} & \text{Core}  & \text{1} \\
    5.0   & 125 \pm 12   & \text{VLBA} & \text{Core}  & \text{1} \\
    15.5  & 80 \pm 11    & \text{VLBA} & \text{Core}  & \text{1} \\
    15.5  & 93 \pm 12    & \text{VLBA} & \text{Core}  & \text{1} \\
    22.0  & 86 \pm 11  & \text{VLBA} & \text{Core}  & \text{1} \\
    22.2  & 101 \pm 11  & \text{EAVN} & \text{Core}  & \text{1} \\
    23.6  & 86 \pm 12    & \text{VLBA} & \text{Core}  & \text{1} \\
    44.1  & 86 \pm 12    & \text{VLBA} & \text{Core}  & \text{1} \\
    87.8  & 83 \pm 16    & \text{VLBA} & \text{Core}  & \text{1} \\
    224.5 & 134 \pm 24   & \text{SMA}  & \text{Core}  & \text{2} \\
    \enddata
    \tablerefs{1.\ \citet{M84Flux1}; 2.\ \citet{M84Flux2}}
\end{deluxetable}

\subsection{NGC 3894}

NGC3894 is an elliptical galaxy that is located 48.2~Mpc away from Earth, as determined using the TFR method \citep{NGC3894Distance}. The mass of the central black hole is estimated as $2 \times 10^9 M_\odot$ from the $M-\sigma$ relation \citep{NGC3894Flux1}. Earliest VLBI studies detected asymmetric structure in radio emission associated with jets \citep{NGC3894EarlyVLBI}. More recent investigation by \citet{NGC3894recentVLBI} estimated a small jet inclination ($10^{\circ}-20^{\circ})$. A helpful compilation of radio and optical/UV observations coupled with an analysis of X-ray and $\gamma$-ray data can be found in \citet{NGC3894Flux1}.  We show in Table~\ref{tab:NGC3894} existing observations in the radio bands.

\begin{deluxetable}{CCCCC}[h]
    \tablecaption{Radio observations of NGC 3894. \label{tab:NGC3894}}
    \tablehead{
        \colhead{$\nu$ (GHz)} & \colhead{Flux (mJy)} & \colhead{Telescope} & \colhead{Region} & \colhead{Reference}
    }
    \startdata
    1.4   & 471.6  $\pm$ 47.2  & \text{VLA}  & \text{Total} & \text{1}  \\
    1.4   & 481.4  $\pm$ 14.4  & \text{VLA}  & \text{Total} & \text{1}  \\
    2.7   & 440    $\pm$ 44    & \text{NRAO*}  & \text{Total} & \text{1}  \\
    2.7   & 440    $\pm$ 11    & \text{RTE**}  & \text{Total} & \text{1}  \\
    4.85  & 627    $\pm$ 62    & \text{GBT}  & \text{Total} & \text{1}  \\
    5     & 660    $\pm$ 20    & \text{VLA}  & \text{Total} & \text{1}  \\
    5     & 110    $\pm$ 16    & \text{VLA} & \text{Core} & \text{2}  \\
    5     & 87     $\pm$ 16    & \text{VLA} & \text{Core} & \text{2}  \\
    5     & 79     $\pm$ 16    & \text{VLA} & \text{Core} & \text{2}  \\
    5     & 67     $\pm$ 16    & \text{VLA} & \text{Core} & \text{2}  \\
    8     & 116    $\pm$ 16    & \text{VLBA} & \text{Core} & \text{2}  \\
    8.1   & 520    $\pm$ 52    & \text{NRAO*}  & \text{Total} & \text{1}  \\
    8.4   & 521.7  $\pm$ 52    & \text{CRATEAS}  & \text{Total} & \text{1}  \\
    8.5   & 436    $\pm$ 4     & \text{VLA}  & \text{Total} & \text{1}  \\
    10.7  & 600    $\pm$ 48    & \text{RTE}  & \text{Total} & \text{1}  \\
    15    & 441    $\pm$ 2     & \text{OVRO 40-m}  & \text{Total} & \text{1}  \\
    22    & 390    $\pm$ 20    & \text{KVN}  & \text{Total} & \text{1}  \\
    30    & 362    $\pm$ 22    & \text{OCRA-p}  & \text{Total} & \text{1}  \\
    90    & 160    $\pm$ 20    & \text{}  & \text{Total} & \text{1}  \\
    230   & 57.6   $\pm$ 4     & \text{SCUBA} & \text{Total} & \text{3}  \\
    \enddata
    \tablerefs{1.\citet{NGC3894Flux1}; 2.\ \citet{NGC3894Flux2}; 3.\ \citet{NGC3894Flux3}}
    \tablecomments{*NRAO 3-element interferometer, **Radio Telescope Effelsberg}
\end{deluxetable}

\subsection{3C 317}

3C~317 is a well-studied radio source within the large elliptical galaxy UGC~9799. The central, supermassive black hole is estimated to have a mass of $4.6 \pm 0.3\times10^9 \, M_{\odot}$ \citep{3C317Mass} and is located 122~Mpc away from Earth \citep{CosmicFlows2}. \citet{3C317Jet} estimates an inclination angle for the jets in the range $81-85^\circ$. For a good review of the source properties and a compilation of radio observations, see \citet{3C317I}. Table~\ref{tab:3C317} lists observations of this source across the radio spectrum.

\begin{deluxetable}{CCCCC}[h]
    \tablecaption{Radio observations of 3C 317 \label{tab:3C317}}
    \tablehead{
        \colhead{$\nu$ (GHz)} & \colhead{Flux (mJy)} & \colhead{Telescope} & \colhead{Region} & \colhead{Reference}
    }
    \startdata
    1.66   & 341  $\pm$ 10.23  & \text{VLA}  & \text{Core}  & \text{1}  \\
    1.66   & 349  $\pm$ 10.47  & \text{VLA}  & \text{Total} & \text{1}  \\
    2.3    & 319               & \text{PTI} & \text{Total} & \text{1} \\
    4.99   & 355  $\pm$ 10.65  & \text{VLA}  & \text{Core}  & \text{1}  \\
    4.99   & 359  $\pm$ 10.77  & \text{VLA}  & \text{Total} & \text{1}  \\
    4.99   & 296  $\pm$ 8.88   & \text{VLA}  & \text{Core}  & \text{1}  \\
    4.99   & 360  $\pm$ 10.8   & \text{VLA}  & \text{Total} & \text{1}  \\
    91.5   & 55   $\pm$ 3      & \text{ALMA} & \text{Core}  & \text{2}  \\
    103.5  & 49   $\pm$ 2      & \text{ALMA} & \text{Core}  & \text{2}  \\
    233    & 34   $\pm$ 2      & \text{ALMA} & \text{Core}  & \text{2}  \\
    343.5  & 43   $\pm$ 4      & \text{ALMA} & \text{Core}  & \text{2}  \\
    \enddata
    \tablerefs{1.\ \citet{3C317I}; 2.\ ALMA Calibrator Catalog}
\end{deluxetable}

\subsection{NGC 4552}

NGC~4552 is an elliptical galaxy that is located 16.3 Mpc away from Earth \citep{NGC4552Distance}, with a central supermassive black hole of $4.9^{+0.7}_{-0.4} \times 10^8 \; M_{\odot}$ \citep{msigma2016}. The nucleus of NGC~4552 was studied over a year-long period by the SMA at 230~GHz; the average and standard deviation of these observations is reported in Table~\ref{tab:NGC4552}, along with some additional, lower-frequency observations.

\begin{deluxetable}{CCCCC}[h]
    \tablecaption{Radio observations of NGC4552 \label{tab:NGC4552}}
    \tablehead{
        \colhead{$\nu$ (GHz)} & \colhead{Flux (mJy)} & \colhead{Telescope} & \colhead{Region} & \colhead{Reference}
    }
    \startdata
    5    & 99.5  & \text{VLBA}  & \text{Core}  & \text{1}  \\
    15   & 58.6  & \text{VLA}   & \text{Core} & \text{1}  \\
    230  & 23 $\pm$ 5 & \text{SMA}  & \text{Core}  & \text{2}  \\
    \enddata
    \tablerefs{1.\ \citet{NGC4552Flux1}; 2.\ \citet{M84Flux2}}
\end{deluxetable}

\subsection{NGC 315}

NGC~315 is a large elliptical galaxy located approximately $70 \; \text{Mpc}$ away from Earth \citep{NGC315NGC4261MassandDistance}, with a central radio source that hosts extensive jets. Recent, high-resolution global VLBI observations estimated a jet viewing angle of $\sim 50^\circ$ and discovered limb-brightening in the jet at parsec scales \citep{NGC315Jet1}. The central black hole has a mass of $2.08^{+0.33}_{-1.4} \times10^9 \, M_{\odot}$ measured from CO gas kinematics \citep{NGC315NGC4261MassandDistance}. Table~\ref{tab:ngc315} lists observations of this source across the radio spectrum.

\begin{deluxetable}{CCCCC}[h]
    \tablecaption{Radio observations of NGC 315 \label{tab:ngc315}}
    \tablehead{
        \colhead{$\nu$ (GHz)} & \colhead{Flux (mJy)} & \colhead{Telescope} & \colhead{Region} & \colhead{Reference}
    }
    \startdata
    5      & 668 $\pm$ 27  & \text{VLA}  & \text{Core}  & \text{1}  \\
    91.5   & 427 $\pm$ 22  & \text{ALMA}  & \text{Total} & \text{2}  \\
    103.5  & 383 $\pm$ 26  & \text{ALMA}  & \text{Total} & \text{2}  \\
    233    & 182 $\pm$ 9   & \text{ALMA}  & \text{Total} & \text{2}  \\
    343.5  & 193 $\pm$ 10  & \text{ALMA}  & \text{Total} & \text{2}  \\
    \enddata
    \tablerefs{1.\ ALMA Calibrator Catalog; 2.\ \citet{NGC315Flux1}}
\end{deluxetable}

\begin{figure}
    \centering
    \includegraphics[width=0.95\linewidth]{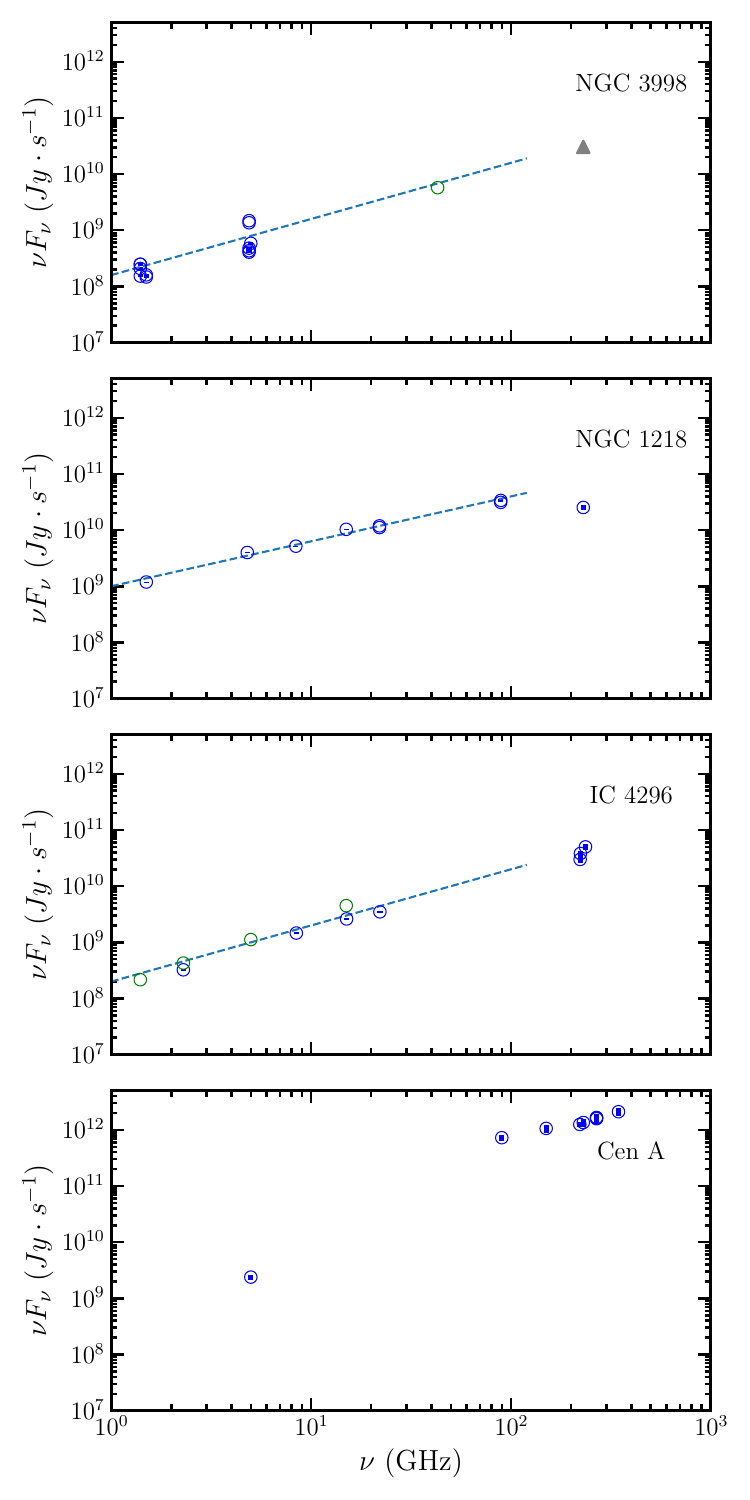}
    \caption{Same as in Fig.~\ref{fig:spectraset1} but for NGC~3998, NGC~1218, IC~4296, and Cen~A. The 230~GHz flux of these sources are not well modeled by a radiatively inefficient accretion flow.}
    \label{fig:enter-label}
\end{figure}

\subsection{NGC 1218}

NGC~1218 is a lenticular galaxy that hosts the radio source 3C78, an active galactic nucleus. Early VLBA studies \citep{NGC1218Flux4,NGC1218Jet1} detected a one-sided jet with an inclination that was determined to be $\sim 20^\circ$ \citep{NGC1218Flux1}. The distance of the galaxy was estimated as $116 \; \text{Mpc}$ \citep{NGC1218Distance}. A recent multi-frequency analysis including observations in the radio band with the VLA and ALMA, in optical wavelengths with the Hubble Space Telescope, in X-rays with Chandra, and in $\gamma$-rays with Fermi-LAT helped constrain the properties of the jet such as the opening angle and demonstrated bulk acceleration in the jet in between pc and kpc scales. For this analysis and a helpful review of the history of observations of this radio source, see \citet{NGC1218Flux1}. Table~\ref{tab:NGC1218} lists observations of this source across the radio spectrum.

\begin{deluxetable}{CCCCC}[h]
    \tablecaption{Radio observations of NGC 1218. \label{tab:NGC1218}}
    \tablehead{
        \colhead{$\nu$ (GHz)} & \colhead{Flux (mJy)} & \colhead{Telescope} & \colhead{Region} & \colhead{Reference}
    }
    \startdata
    1.5   & 794 $\pm$ 3    & \text{VLA}   & \text{Core}  & 1 \\
    1.5   & 752            &  \text{VLA}     & \text{Core} & 4 \\
    4.8   & 832 $\pm$ 25   & \text{VLA}   & \text{Core}  &  1\\
    5     & 628            &  \text{VLA}   & \text{Core} & 4 \\
    8.4   & 616 $\pm$ 1    & \text{VLA}   & \text{Core}  & 1 \\
    15    & 689 $\pm$ 3    & \text{VLA}   & \text{Core}  & 1 \\
    15    & 689            & \text{VLA}      & \text{Core} & 4 \\
    22    & 504            & \text{VLA}   & \text{Core}  &  1\\
    22    & 541            & \text{VLA}   & \text{Core}  &  1\\
    88.9  & 351            & \text{ALMA}  & \text{Core}  &  1\\
    89    & 380 $\pm$ 20   & \text{IRAM}  & \text{Core}  &  3\\
    230   & 110 $\pm$ 10   & \text{IRAM}  & \text{Core}  &  3\\
    345   & 280            & \text{HHT}       & \text{Core} &  2\\
    \enddata
    \tablerefs{1.\ \citet{NGC1218Flux1}; 2.\ \citet{NGC1218Flux2}; 3.\ \citet{NGC1218Flux3}; 4.\ \citet{NGC1218Flux4}}
\end{deluxetable}

\subsection{IC 4296}

IC~4296 is a nearby elliptical galaxy with a radio core hosting relatively symmetric dual jets as viewed by the VLA \citep{IC4296Jet1}. The mass of the central black hole is estimated to be $ 1.34^{+2.1}_{-1.9} \times 10^9 M_{\odot}$ from gas-dynamical models \citep{IC4296Mass} and its distance to be $50.8\;$Mpc, determined from infrared SBF methods \citep{IC4296Distance}. Some studies have suggested difficulty with matching the emission of the core to a radiatively inefficient model \citep{IC4296Flux1}, while others have suggested that a model including outflows would explain the discrepancies with the mass accretion rate \citep{SMAFluxSurvey}. In our analysis, we adopt the flux measured by ALMA (see Table \ref{tab:IC4296}).

\begin{deluxetable}{CCCCC}[t]
    \tablecaption{Radio observations of IC 4296\label{tab:IC4296}}
    \tablehead{
        \colhead{$\nu$ (GHz)} & \colhead{Flux (mJy)} & \colhead{Telescope} & \colhead{Region} & \colhead{Reference}
    }
    \startdata
    1.4 & 154 &  \text{VLA} & \text{Core*}  & 1\\
    2.3  & 186    & \text{VLA} & \text{Core*}  &  1\\
    2.3  & 140.7 $\pm$ 5.5    & \text{VLBA} & \text{Core}  & 1 \\
    5  & 224    & \text{VLA} & \text{Core*}  &  \\
    8.4  & 173.2 $\pm$ 6.8    & \text{VLBA} & \text{Core}  &  1\\
    15  & 299.7   & \text{VLA} & \text{Core*}  & 1 \\
    15 & 173.0 $\pm$ 7.1    & \text{VLBA} & \text{Core}  & 1 \\
    22 & 157.6 $\pm$ 6.15   & \text{VLBA} & \text{Core}  &  1\\
    221.8  & 134.9 $\pm$ 18.9   & \text{SMA} & \text{Core}  &  2\\
    222.5  & 171.5 $\pm$ 17.6   & \text{SMA} & \text{Core}  & 2 \\
    236    & 212.1 $\pm$ 22     & \text{ALMA} & \text{Core}  &  3\\
    \enddata
    \tablerefs{1.\ \citet{IC4296Flux1}; 2.\ \citet{SMAFluxSurvey}; 3.\ \citet{IC4296Flux3} }
    \tablecomments{*kpc scale observations}
\end{deluxetable}

\subsection{NGC 1399}

NGC~1399 is a large elliptical galaxy that is located in the Fornax cluster. It hosts a radio source with approximately symmetric radio lobes \citep{NGC1399Flux1}. While there are few radio observations of this source, since it is relatively dim, it has been studied in the mm regime by the SMA \citep{SMAFluxSurvey}. The mass of the central supermassive black hole was estimated to be $5.1 \pm 0.7 \times10^8 \;M_\odot$ using stellar dynamical modeling of  Hubble observations \citep{NGC1399Mass}. The distance to the galaxy is $21.1$ Mpc, estimated from infrared SBF methods \citep{NGC1399Distance}. Table~\ref{tab:NGC1399} lists the available radio observations.

\begin{deluxetable}{CCCCC}[h]
    \tablecaption{Radio observations of NGC 1399\label{tab:NGC1399}}
    \tablehead{
        \colhead{$\nu$ (GHz)} & \colhead{Flux (mJy)} & \colhead{Telescope} & \colhead{Region} & \colhead{Reference}
    }
    \startdata
    1.4 & 463 \pm 46.3 &  \text{VLA} & \text{Total}  & 1\\
    221.7  & 38.5 \pm 6.4    & \text{SMA} & \text{Core}  &  2\\
    \enddata
    \tablerefs{1.\ \citet{NGC1399Flux1}; 2.\ \cite{SMAFluxSurvey}}
\end{deluxetable}

\subsection{Cen A}

Centaurus~A is the closest galaxy in this sample, with a distance of only 3.42~Mpc, determined using Cepheids \citep{CenADistance}. Due to its relatively small black hole mass, $5.5 \pm 0.3 \times 10^7 \;M_\odot$, as measured from stellar kinematics \citep{CenAMass}, its projected shadow diameter lies just above the 1.5~$\mu$as cut we have imposed for shadow imaging. However, its proximity, jets, and brightness in the radio band led us to keep this source for potential observation by advanced radio interferometry missions. Cen~A has been observed by the Event Horizon Telescope, resolving the jet collimation near the central black hole at hundreds of gravitational radii \citep{EHTCenA}. Additional observations are listed in Table~\ref{tab:CenA}.

\begin{deluxetable}{CCCCC}[h]
    \tablecaption{Radio observations of Cen A\label{tab:CenA}}
    \tablehead{
        \colhead{$\nu$ (GHz)} & \colhead{Flux (mJy)} & \colhead{Telescope} & \colhead{Region} & \colhead{Reference}
    }
    \startdata
    \
    90    & 8063.3 $\pm$ 1059.8 & \text{JCMT*}  & \text{Core}  & \text{1} \\
    150   & 7088.0 $\pm$ 1132.2 & \text{JCMT}  & \text{Core}  & \text{1} \\
    221   & 5660.0 $\pm$ 570.0  & \text{ALMA}  & \text{Core}  & \text{2} \\
    230   & 5877.5 $\pm$ 941.2  & \text{JCMT}  & \text{Core}  & \text{1} \\
    268   & 5900.0 $\pm$ 900.0  & \text{JCMT}  & \text{Core}  & \text{1} \\
    268   & 6100.0 $\pm$ 900.0  & \text{JCMT}  & \text{Core}  & \text{1} \\
    268   & 6200.0 $\pm$ 900.0  & \text{JCMT}  & \text{Core}  & \text{1} \\
    345   & 6100.0 $\pm$ 1011.9 & \text{JCMT}  & \text{Core}  & \text{1} \\
    \enddata
    \tablerefs{1. \citet{CenAFlux1} 2.\citet{PolarimetricMeasureofEHTTarget}}
    \tablecomments{*James Clerk Maxwell Telescope (JCMT)}
\end{deluxetable}

\section{Scaling of the Transition Frequency}
\label{sec: Appendix B}

In this Appendix, we present analytical arguments and semi-analytic calculations to determine the frequency at which a radiatively inefficient accretion flow becomes optically thin and to infer the scaling of this transition frequency with black hole mass and accretion rate.

We start with the radiative transfer equation at frequency $\nu$ 
\begin{equation}
    \frac{dI_{\nu}}{ds} = j_\nu - \alpha_{\nu} I_{\nu} \;,
\end{equation}
where $I_{\nu}$ is the specific intensity, $j_\nu$ is the emissivity, $\alpha_\nu$ is the absorption coefficient, and $s$ is the path length through the material. By rearranging this equation and expressing it in terms of the optical depth, 
\begin{equation}\label{eq: tau_integral}
    \tau_\nu = \int \alpha_\nu ds
\end{equation}
we obtain
\begin{equation}
    \frac{dI_{\nu}}{d\tau_\nu} = S_\nu - I_\nu\;,
\end{equation}
where $S_\nu = j_\nu/\alpha_\nu$ is the source function. Assuming a thermal population of electrons and using Kirchhoff's law, the absorption coefficient $\alpha_\nu$ is related to the emissivity by $\alpha_\nu = j_\nu/B_\nu(T_e)$, where $B_\nu(T_e)$ is the blackbody function at the local electron temperature $T_e$.

\begin{figure}[t]
    \centering
    \includegraphics[width=0.99\linewidth]{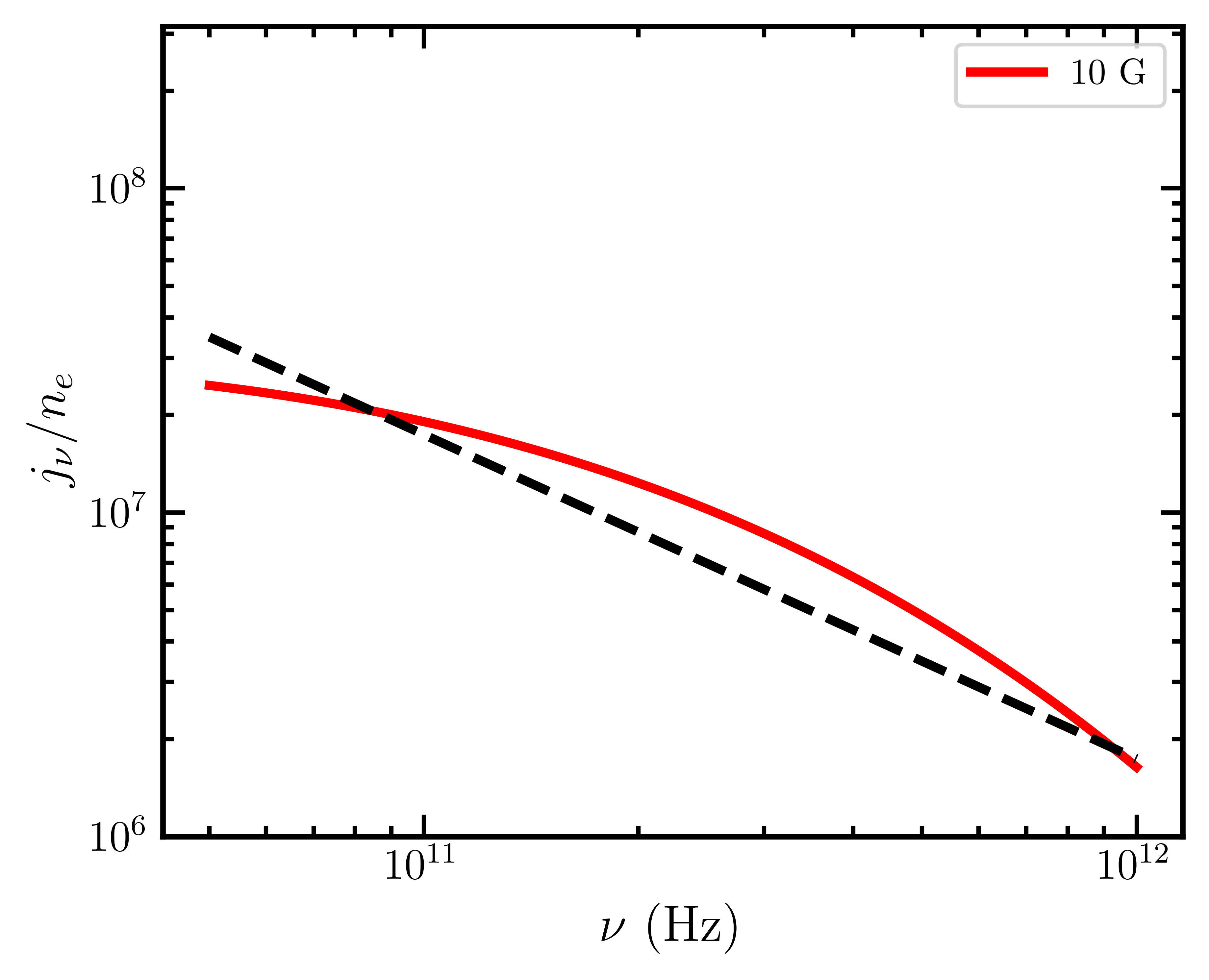}
    \caption{Dependence of the synchrotron emissivity divided by electron density on the observing frequency. The dashed line shows a power-law relation of the form $j_\nu \propto \nu^{-1}$.}
    \label{fig:emissivityvsfreq}
\end{figure}

\begin{figure}[t]
    \centering
    \includegraphics[width=0.99\linewidth]{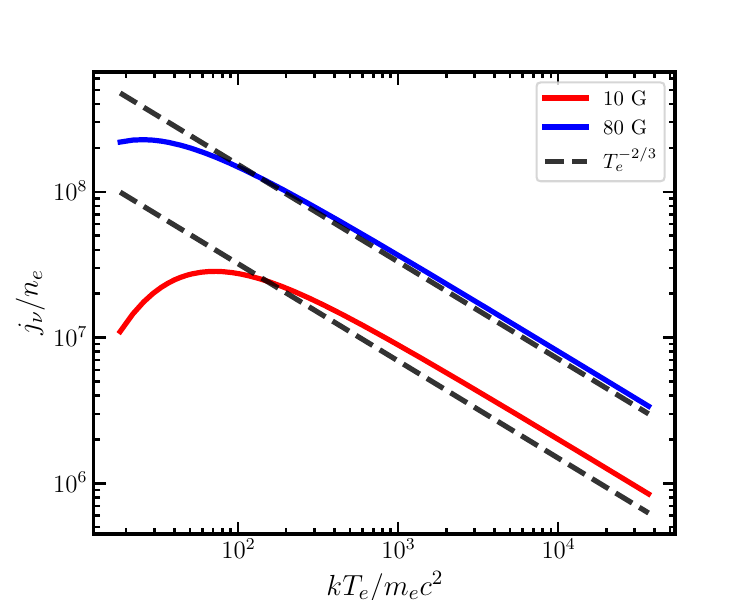}
    \caption{Dependence of the synchrotron emissivity divided by electron density on the electron temperature for two values of the magnetic field strength. The dashed lines show power-law relations of the form $j_\nu \propto T_e^{-2/3}$.}
    \label{fig:emissivityvsT}
\end{figure}

The transition frequency from optically thick to optically thin emission is the frequency at which the optical depth through the flow becomes $\tau_\nu=1$. \citet{SatapathyVariability} found that the synchrotron emissivity for pertinent conditions of radiatively inefficient accretion disks scales as $j_\nu \propto n_e B^2$. The emissivity also has a complex dependence on temperature and on the frequency of observation; yet, in the conditions of interest for density, temperature, and magnetic field, we find that the synchrotron emissivity scales approximately as $j_\nu \propto \nu^{-1} T_e^{-2/3}$ (see Figs.  \ref{fig:emissivityvsfreq} and  \ref{fig:emissivityvsT}). 

In radiatively inefficient accretion flows, the plasma~$\beta$ parameter is always of order $5-10$ at horizon scales, suggesting that the square of the magnetic field scales as the thermal pressure, i.e, $B^2\propto n_e T_e$. Combined with the above scalings, we find that the synchrotron emissivity scales as $j_\nu \propto n_e^2 T_e^{1/3}\nu^{-1}$.

\begin{figure}[t]
    \centering
    \includegraphics[width=0.99\linewidth]{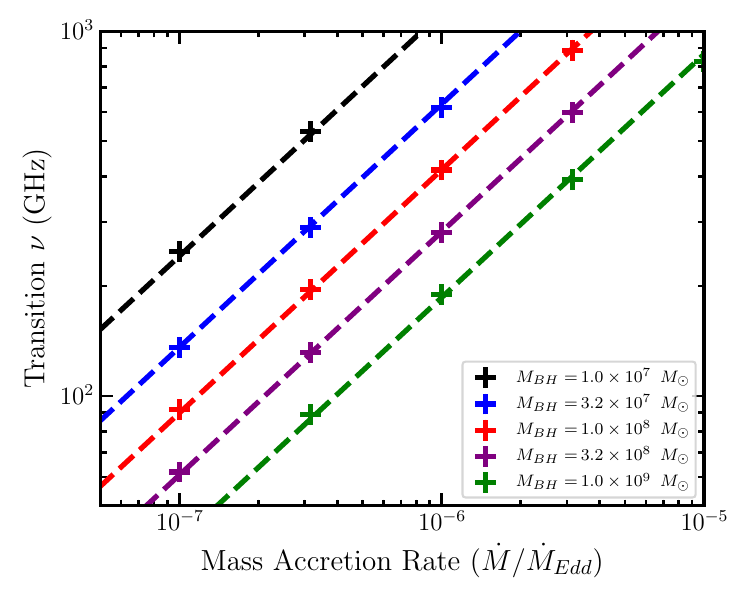}
     \caption{Transition frequencies from optically thick to optically thin emission in radiatively inefficient accretion flows, as a function of mass accretion rate, for different black hole masses. The remaining parameters of the accretion flow model are $a = 0.5$, $i=30^\circ$, $n_r=1.5$, $\beta = 10$, $\zeta=0.4$,  Dashed lines show power-law relations of the form $\nu\propto (\dot{M}/\dot{M}_{\rm Edd})^{2/3}$.\label{fig:transitionnuvsmdot}}
\end{figure}

The electron density at coordinate radius $r$ is related to the geometry of the infalling material and the mass accretion rate $\dot{M}$ through mass conservation, i.e., 
\begin{equation} 
\label{eq: massaccretion}
    n_e = \frac{\dot{M}}{4\pi (h/r) r^2 u^r m_p } \;,
\end{equation}
where $h/r$ is the scale height of the flow, $u^r$ is the radial velocity, and $m_p$ is the proton mass. Expressing all radii in gravitational units, the mass accretion rate in terms of the Eddington rate as $\dot{M}= \dot{m} \dot{M}_{\rm Edd}$, and writing the radial velocity as a power-law function of radius (see eq.~[\ref{eq:radial velocity profile}]), 
we obtain
\begin{equation}
\label{eq: density scaling}
    n_e \propto \frac{\dot{m}}{M}r^{-2+n_r}.
\end{equation}

In the radio frequencies of interest, the blackbody function can be approximated with the Rayleigh-Jeans law $B_{\nu} = 2 k_b T_e \nu^2/c^2$. Incorporating these scalings into equation~(\ref{eq: tau_integral}), we obtain
\begin{equation}
\label{eq: taupropto}
    \tau \propto\int_{r_{in}} \frac{n_e^2 T_e^{1/3}}{\nu} \frac{1}{\nu^2 T_e}ds
    = \frac{1}{\nu^3}\int_{r_{in}} \frac{n_e^2 }{ T_e^{2/3}}ds\;.
\end{equation}

\begin{figure}[t]
    \centering
    \includegraphics[width=0.99\linewidth]{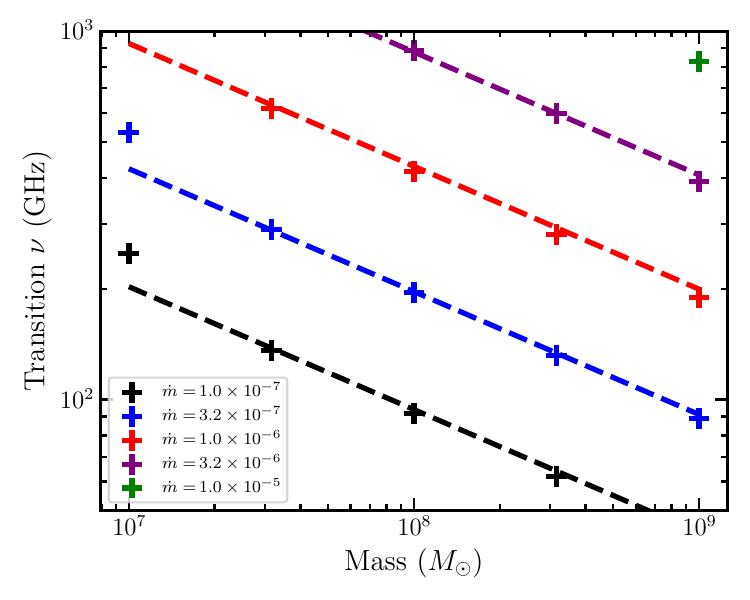}
    \caption{Same as Figure~\ref{fig:transitionnuvsmdot} but showing the dependence of the transition frequency on black-hole mass, for different accretion rates. Dashed lines show power-law relations of the form $\nu\propto M^{-1}$.}
    \label{fig:transitionnuvsmass}
\end{figure}

Using equation~(\ref{eq:Te}) for the radial dependence of the electron temperature, expressing the path length in geometric units, and extending the integral to the black hole horizon, we then find
\begin{equation}
\label{eq: taupropto}
    \tau \propto \frac{\dot{m}^2}{\nu^3M}\int_{r_{\rm hor}} r^{-10/3+2n_r} ds \;.
\end{equation}
The integral in this expression evaluates to a number, leading to
\begin{equation}
\label{eq: tau_scaling}
    \tau \propto \frac{\dot{m}^2}{M\nu^3}\;.   
\end{equation}

We test this frequency scaling by performing radiative transfer calculations of accretion flows varying black hole mass and accretion rate and evaluating numerically the transition frequency in each case. We show the dependence on mass accretion rate in Figure~\ref{fig:transitionnuvsmdot} and on black hole mass in Figure~\ref{fig:transitionnuvsmass}, both of which closely follow the analytic scaling derived here.

\bibliography{source}{}
\bibliographystyle{aasjournal}

\end{document}